\documentclass[10pt,twocolumn,showpacs,longbibliography,amsmath,amssymb,osajnl,floatfix,superscriptaddress]{revtex4-1}

\usepackage{graphicx}
\usepackage{dcolumn}
\usepackage{bm}
\usepackage{color}
\usepackage{txfonts}
\usepackage{microtype}

\begin{document}
\author{Guang Hu}
\affiliation{Department of Nuclear Science and Technology, Xi'an Jiaotong University, Xi'an 710049, China}
\author{Wei-Qiang Sun}
\affiliation{Department of Nuclear Science and Technology, Xi'an Jiaotong University, Xi'an 710049, China}		
\author{Bing-Jun Li}
\affiliation{Department of Nuclear Science and Technology, Xi'an Jiaotong University, Xi'an 710049, China}
\author{Yan-Fei Li}\email{liyanfei@xjtu.edu.cn}	
\affiliation{Department of Nuclear Science and Technology, Xi'an Jiaotong University, Xi'an 710049, China}
\author{Wei-Min Wang}\email{weiminwang1@ruc.edu.cn}
\affiliation{Department of Physics and Beijing Key Laboratory of
Opto-electronic Functional Materials and Micro-nano Devices, Renmin
University of China, Beijing 100872, China} \affiliation{Beijing National Laboratory for 
Condensed Matter Physics, Institute of Physics, CAS, Beijing 100190,
China}\affiliation{Collaborative Innovation Center of IFSA (CICIFSA), Shanghai Jiao Tong University, Shanghai 200240, China}
\author{Meng Zhu}
\affiliation{Northwest Institute of Nuclear Technology, Xi'an 710024, China}		
\author{Hua-Si Hu}\email{huasi\_hu@mail.xjtu.edu.cn}	
\affiliation{Department of Nuclear Science and Technology, Xi'an Jiaotong University, Xi'an 710049, China}
\author{Yu-Tong Li}
 \affiliation{Beijing National Laboratory for 
Condensed Matter Physics, Institute of Physics, CAS, Beijing 100190,
China}\affiliation{School of Physical Sciences, University of Chinese Academy of Sciences, Beijing 100049, China}
\affiliation{Collaborative Innovation Center of IFSA (CICIFSA), Shanghai Jiao Tong University, Shanghai 200240, China}
\affiliation{Songshan Lake Materials Laboratory, Dongguan, Guangdong 523808, China}

\title{Quantum-stochasticity-induced asymmetry in angular distribution of electrons in a quasi-classical regime}

\date{\today}

\begin{abstract}

Impacts of quantum stochasticity on the dynamics of an ultra-relativistic electron beam head-on colliding with a linearly polarized ultra-intense laser pulse are theoretically investigated in a quasi-classical regime. 
Generally, the angular distribution of the electron beam keeps  symmetrically in transverse directions in this regime, even under the ponderomotive force of the laser pulse.  Here we show that when the initial angular divergence $\Delta \theta_i\lesssim 10^{-6}a_0^2$ with $a_0$ being the normalized laser field amplitude, an asymmetric angular distribution of the electron beam arises due to the quantum stochasticity effect, via simulations employing Landau-Lifshitz, quantum-modified Landau-Lifshitz equations,  and quantum stochastic radiation reaction form to describe the radiative electron dynamics respectively. 
The asymmetry is robust against a variety of laser and electron parameters, providing an experimentally detectable signature for the nature of quantum stochasticity of photon emission with laser and electron beams currently available.

\end{abstract}

\maketitle
Dynamics of an electron in electromagnetic fields is a fundamental issue in either classical electrodynamics \cite{Jackson1998} or quantum electrodynamics (QED) \cite{Berestetskii1982}. Apart from the main Lorentz force, the electron also suffers from the reaction force of radiation. In classical realm, the radiation reaction (RR) effect is taken as radiation damping stemming from the radiated electromagnetic fields coupling the external fields. The well-known Lorentz-Abraham-Dirac (LAD) equation \cite{Abraham1905,Lorentz1909,Dirac1938} self-consistently describes the electron motion accounting for RR effects as an additional four-force. However, LAD equation gives unphysical solutions, such as the ``runaway'' solution. To fix it, Landau-Lifshitz (LL) equation, was developed through bringing a perturbative iteration in the RR terms in the LAD equation, and it is employed as the classical equation of electron motion at relatively low electromagnetic wave amplitude \cite{Landau1975}. On the other hand, LL equation overestimates the radiative energy loss, since it unphysically includes the emission of photons with energy higher than the electron kinetic energy \cite{Yoffe2015}. Recently, a quantum-modified LL equation is derived with quantum-recoil corrections through rescaling the RR force by a factor of $I_{QED}/I_{C}$ (i.e. the ratio of the radiation intensities within QED and classical approaches) \cite{Sokolov2010,Erber1966}, avoiding the aforementioned classical overestimation. Another distinguishing property of radiation in QED is the nature of quantum stochasticity, i.e., the discrete and probabilistic character of photon emission \cite{Neitz2013,Neitz2014,Wan2019}. The quantum stochasticity effect (QSE) would increase the yield of high-energy photons \cite{Blackburn2014}, cause the quantum quenching of radiation losses \cite{Harvey2017}, alter the energy spectrum of emitted photons \cite{Wan2019} or scattered electrons \cite{Neitz2013,Piazza2009,Niel2018}, reshape the space-distribution of photons \cite{Li2017} or electrons \cite{Liyf2018,Baird2019}, etc \cite{Bashinov2015,Wang2015,Li2020}. 

Nowadays, the development of ultra-short ultra-intense laser techniques, particularly the application of PW lasers \cite{Danson2019}, has stimulated the research interests in confirmatory experiments on QED theory \cite{Piazza2012,Cole2018,Poder2018}. Recently, quantum RR was reported to be observed in experiments of laser-electron-beam interaction \cite{Cole2018,Poder2018}, via comparing the electron- and photon-spectra detected with those simulated in quantum theoretical model. Note that in these experiments, the QSE is detected mixed with other quantum properties, such as quantum recoil. 

Proposals \cite{Liyf2018,Neitz2013,Harvey2017,Blackburn2014,Neitz2014,Wan2019,Li2017,Piazza2009,Niel2018} for unambiguous identification of QSE mainly involves quantum radiation dominated regime (QRDR) characterized by the parameter of $R_c=\alpha \chi a_0 \gtrsim1$ \cite{Koga2005,Piazza2012}. Here, $\alpha\approx1/137$ is the fine-structure constant, indicating the order of photon-emission probability of the electron in a formation length ($l_f\sim \lambda_0/a_0$ with $\lambda_0$ being the laser wavelength \cite{Ritus1985}); $a_0 \equiv eE_0/(m\omega_0) $ is the normalized laser field parameter, $\omega_0$ the laser frequency, and $e (>0),m$ the electron charge and mass, respectively; and $\chi \equiv e \sqrt{-(F_{\mu v}p^v)^2}/m^3\approx 2 \omega_0 a_0 \gamma/m$ is the nonlinear electron quantum parameter corresponding to the ratio of the typical emitted photon energy with the initial electron kinetic energy as $\omega_\gamma/\varepsilon_0\sim \chi$, where $F_{\mu v}$ is the field tensor and $p^v$ the is four-vector of electron momentum. Units $\hbar=c=1$ are used throughout. Apparently, the QRDR represents a regime where the energy loss of an electron in a laser period due to RR is comparable with its initial energy. The QRDR can elicit remarkable impacts of the RR effects on electron dynamics even when $\chi \ll 1$, whereas it also requires super high electron energy and laser intensity in experiments.  Moreover, signatures associated with electron- and/or photon-spectrum \cite{Arran2019,Neitz2013,Piazza2009,Niel2018,Blackburn2014,Harvey2017} would be submerged by the fluctuation and statistical uncertainty of the laser and electron beam parameters. Even though in some measurement methods the experimental feasibility for discerning the QSE has been proved numerically \cite{Baird2019,Liyf2018,Arran2019}, the fact that QSE has not been observed distinguishably promotes more investigations on QSE signatures.

To tap the signature of QSE in regime of  $R_c\ll1$ with significantly reduced laser intensity and electron energy, we investigate the dynamics of an ultra-relativistic electron beam in nonlinear Compton scattering process with a linearly polarized ultra-intense laser pulse field in a quasi-classical regime with $\chi\ll1$. With QSE, an asymmetric radiation energy loss arising from the discrete and probabilistic character of photon emission can lead to a notable deflection effect on the scatted electrons, thus imprint a laser-polarization-dependent asymmetry on the electron distribution. Without QSE, the RR effects on electron dynamics which can be considered as damping force proportional to $E_0^2$ and opposite to its velocity, would result in a negligible integrated deflection effect on these electrons. Simulations based on LL equation (LLM), modified-LL equation (MLLM) and Monte-Carlo stochastic model (MCM), respectively, to describe the electron dynamics including RR effects, are performed for qualitative and quantitative studies.

\begin{figure}[t]
 	\includegraphics[width=0.9\linewidth]{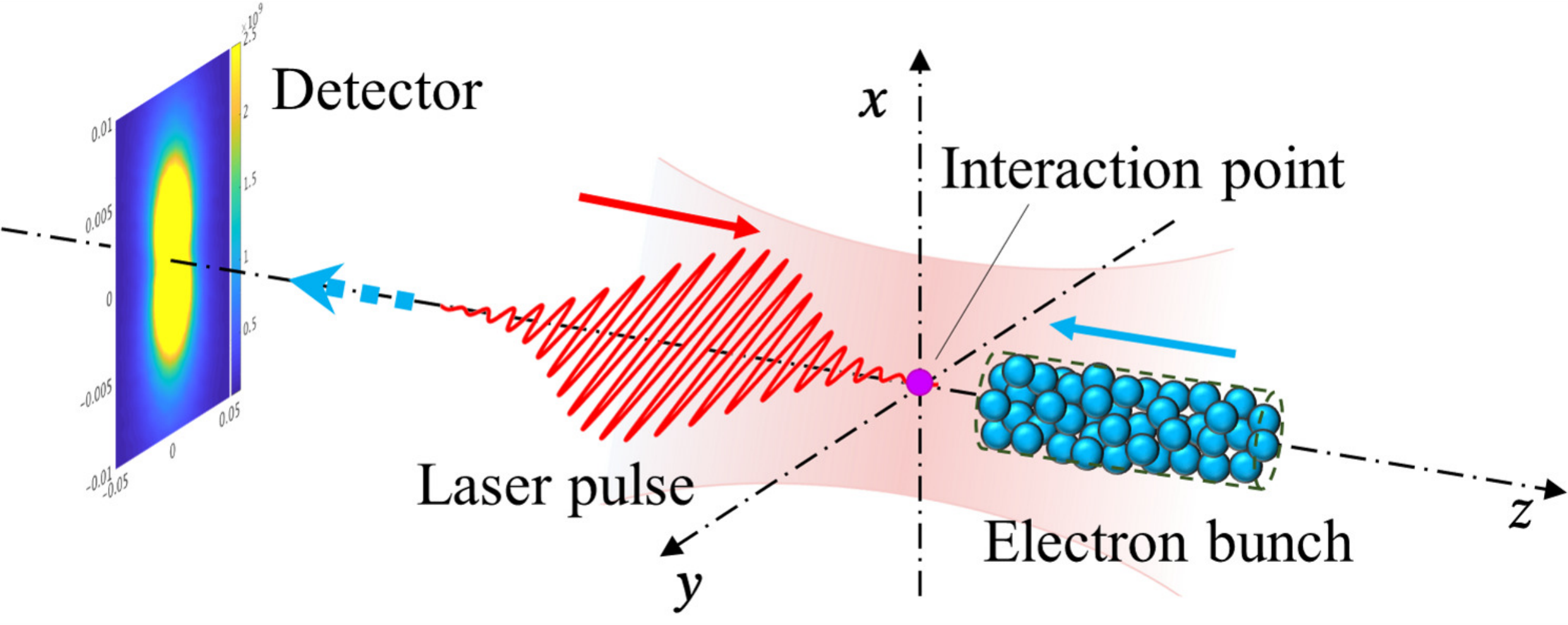}
\caption{Scenario of laser-electron-beam interaction to reveal the QSE on electron-beam dynamics. The electron bunch  propagating along $-z$ head-on collides with a linearly polarized ultra-intense laser pulse. Signature of QSE could be observed from the detectors recoding the final angular distribution of the electrons.} \label{fig1}
\end{figure}

The aforementioned simulation methods would be introduced briefly below, with details described in our previous work \cite{Liyf2018} or Refs.\cite{Piazza2012,Poder2018}. The LLM is a pure classical method with the tree-dimensional RR force reading \cite{Landau1975}:
 \begin{eqnarray}\label{LL}
{\bm F_{RR}}&=&-\frac{2e^3}{3mc^3} \left\{\gamma\left[\left(\frac{\partial}{\partial t}+\frac{\textbf{p}}{\gamma m}\cdot\nabla\right){\textbf{E}}+\frac{{\textbf{p}}}{\gamma m c}\times\left(\frac{\partial}{\partial t}+\frac{{\textbf{p}}}{\gamma m}\cdot\nabla\right){\textbf{B}}\right]\right.\nonumber\\
&& -\frac{e}{m c}\left[{\textbf{E}}\times{\textbf{B}}+\frac{1}{\gamma m c}{\textbf{B}}\times\left({\textbf{B}}\times{\textbf{p}}\right)+\frac{1}{\gamma m c}{\textbf{E}}\left({\textbf{p}}\cdot{\textbf{E}}\right)\right]\nonumber\\
&& \left. +\frac{e\gamma}{m^2 c^2}{\textbf{p}}\left[\left({\textbf{E}}+\frac{{\textbf{p}}}{\gamma m c}\times {\textbf{B}}\right)^2-\frac{1}{\gamma^2m^2c^2}\left({\textbf{E}}\cdot{\textbf{p}}\right)^2\right]\right\},
\end{eqnarray}
where $\textbf{E}$ and $\textbf{B}$ are the electric and magnetic fields, respectively. 
Quantum correction to the RR force is conducted in MLLM by adding a modifying factor of $g(\chi)\equiv I_{QED}/I_C$ \cite{Sokolov2010,Erber1966}, i.e., 
\begin{eqnarray}\label{MLL}
{\bm F_{RR}'}&=&g(\chi){\bm F_{RR}},
\end{eqnarray}
where, $I_{QED}=\int \omega_\gamma$ ${\rm d} W_{rad}/({\rm d} t {\rm d} \omega_\gamma) {\rm d} \omega_\gamma$ indicates the quantum total emission power, with $\omega_\gamma$ the emitted photon energy and $W_{rad}$ the radiation probability; and $I_{C}=2e^4E'^2/(3m^2)$ is the corresponding classical quantity calculated at the local value of $E'$, the electric fields in the electron frame. 
Apparently, the LLM and MLLM treat the RR effects excluding QSE.  

The MCM, deals with the photon emission quantum mechanically, fusing the QSE into RR process by taking advantage of Monte-Carlo stochastic algorithm. The discrete and probabilistic photon emission is performed by a stochastic procedure based on the radiation probability in the local constant approximation \cite{Ritus1985,Khokonov2002,Sokolov2010,Piazza2018,Piazza2019,Harvey2015,Green2015,Elkina2011,Ridgers2014}, which reads \cite{Baier1998}:
\begin{eqnarray}\label{W}
{\rm d}W_{rad}&=&\frac{{\alpha m}}{{\sqrt{3}\pi \gamma}}\left[{\rm IntK}_{\frac{5}{3}}(u')+\frac{\delta^2}{1-\delta}{\rm K}_{\frac{2}{3}}(u')\right]{\rm d}\delta{\rm d}t;
\end{eqnarray}
where, $u'=2\delta/[3\chi(1-\delta)]$ with $\delta=\omega_\gamma/\varepsilon_0$, ${\rm IntK}_{\frac{5}{3}}(u')\equiv \int_{u'}^{\infty} {\rm d}z {\rm K}_{\frac{5}{3}}(z)$ with ${\rm K}_n$ being the $n$-order modified Bessel function of the second kind. In each time step $\Delta t$, the probability for an electron to emit a photon with energy of $\omega_\gamma=\delta \varepsilon_i$ ($0<\delta<1$) is calculated with  Eq.(\ref{W}), i.e. $W_{rad}(\delta)={\rm d} W_{rad}/({\rm d}\delta {\rm d}t)\Delta t\Delta \delta$. To avoid an infrared cutoff, we take $\delta=r_1^3$ \cite{Gonoskov2015}, with $r_1$ being a random number in [0,1].  Another random number  $r_2\in [0,1]$ is used to determine if a photon is emitted: if $W_{rad}(r_1)<r_2$, reject emitting a photon; otherwise, emit a photon of $\omega_\gamma$. Given the smallness of the emission angle $1/\gamma$ for an ultra-relativistic electron, the emitted photon is assumed to move along the electron velocity. Here, the $\gamma$ is the electron Lorentz factor. More detailed information on this method and its accuracy have been shown in Ref.\cite{Gonoskov2015}. Between emissions, the electron dynamics in the laser field is governed by classical Lorentz equations of motion.

 In our simulations, effects from electron spin or emitted photon polarization are ignored due to its negligible integrated influence in the nonlinear Compton scattering process \cite{Liyf2019,Liyf2020}.  While similar classical \cite{Burton2014,Vranic2016} and quantum \cite{CAIN,Ridgers2014,Gonoskov2015} simulation models have also been put forward, the signature of QSE, i.e. the differences between the classical and quantum models, are expected to be invariant for the scheme and parameters considered here, as shown below. 

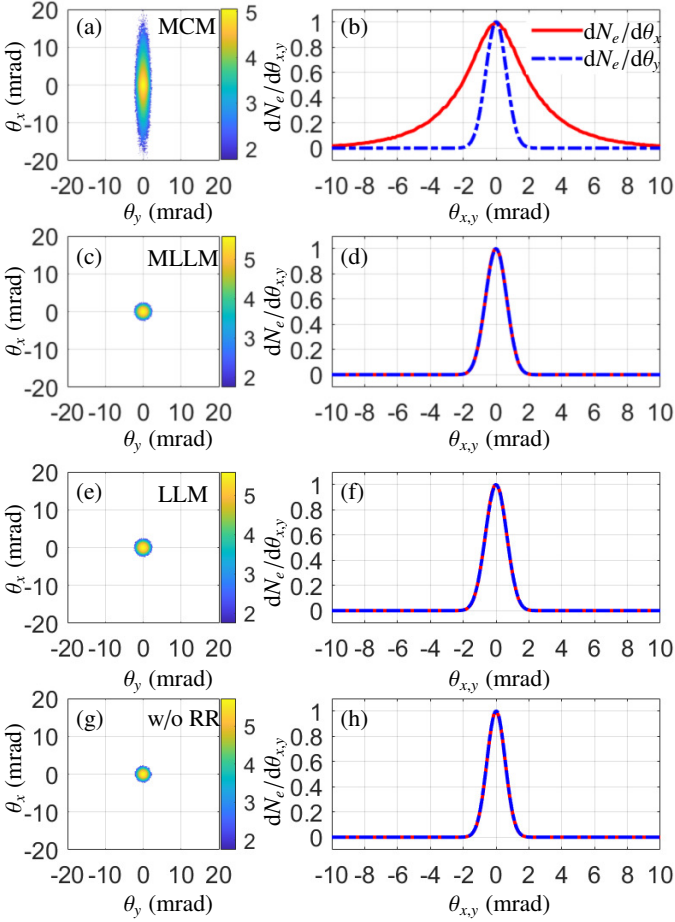
\begin{figure}[t]
 	\includegraphics[width=1\linewidth]{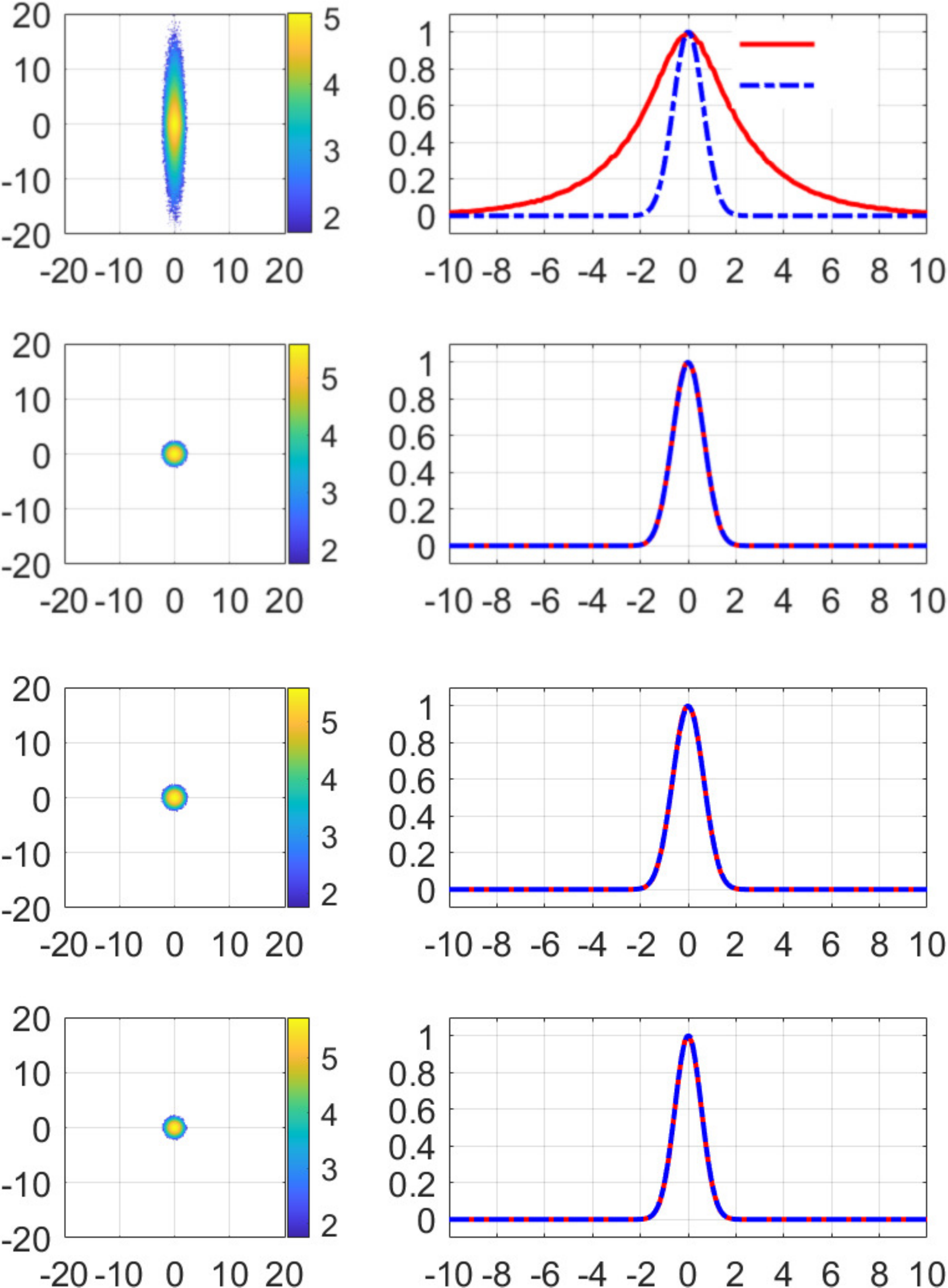}
 \begin{picture}(200,0)
       \put(-3,332){(a)}
       \put(28,332){MCM}
       \put(15,262){\small $\theta_y$ (mrad)}
       \put(-30,295){\rotatebox{90}{\small $\theta_x$ (mrad)}} 
       
       \put(97,332){(b)}
       \put(138,262){\small $\theta_{x,y}$ (mrad)}
       \put(68,295){\rotatebox{90}{\small ${\rm d}N_e/{\rm d}\theta_{x,y}$}} 
       \put(189,330){\small ${\rm d}N_e/{\rm d}\theta_{x}$} 
       \put(189,320){\small ${\rm d}N_e/{\rm d}\theta_{y}$}
       
       \put(-3,245){(c)}
        \put(24,244){MLLM}
       \put(15,176){\small $\theta_y$ (mrad)}
       \put(-30,210){\rotatebox{90}{\small $\theta_x$ (mrad)}}
       
       \put(97,245){(d)}
       \put(138,176){\small $\theta_{x,y}$ (mrad)}
       \put(68,210){\rotatebox{90}{\small ${\rm d}N_e/{\rm d}\theta_{x,y}$}} 
       
       \put(-3,156){(e)}
        \put(28,155){LLM}
       \put(15,85){\small $\theta_y$ (mrad)}
       \put(-30,120){\rotatebox{90}{\small $\theta_x$ (mrad)}}
       
       \put(97,156){(f)}
       \put(138,85){\small $\theta_{x,y}$ (mrad)}
       \put(68,115){\rotatebox{90}{\small ${\rm d}N_e/{\rm d}\theta_{x,y}$}} 
       
       \put(-3,70){(g)}
       \put(24,70){w/o RR}
       \put(15,0){\small $\theta_y$ (mrad)}
       \put(-30,37){\rotatebox{90}{\small $\theta_x$ (mrad)}}
       
       \put(97,70){(h)}
       \put(138,0){\small $\theta_{x,y}$ (mrad)}
       \put(68,32){\rotatebox{90}{\small ${\rm d}N_e/{\rm d}\theta_{x,y}$}}
\end{picture}
\caption{Two-dimension distribution of electron number density of ${\rm d}^2N_e/({\rm d}\theta_x {\rm d}\theta_y)$ (mrad$^{-2}$) ( left column), vs deflection angles of $\theta_x=p_x/p_z$ and $\theta_y=p_y/p_z$; and integrated one-dimension distribution of electron density of ${\rm d}N_e/{\rm d}\theta_{x,y} $ (mrad$^{-1}$) (right column), vs $\theta_x$ (red-solid) or $\theta_y$ (blue-dash-dotted). Rows from top to bottom are the simulated results calculated with RR in MCM, MLLM and LLM, respectively; and without RR.  } \label{fig2}
\end{figure}

A typical simulation result, employing a feasible scenario involving an electron beam of  $\varepsilon_0=300$ MeV and a laser pulse with peak intensity of $a_0=30$ or $I_0=1.2\times 10^{21}$W/cm$^2$ (corresponding to the quantum parameters of $\chi_{max}\approx 0.08$ and $R_c\approx0.02$), is illustrated in Fig.~\ref{fig2}. 
The electron bunch is set with features of laser-accelerated electron
source \cite{Esarey2009,Leemans2014,Gonsalves2019}: $N_e=1\times10^5$ electrons uniformly distributed
longitudinally  in a cylindrical form at length of $L_e=5 \mu$m and normally distributed transversely in a radius of $R_e=1 \mu$m with standard deviation of $\sigma_{x,y}=0.3 \mu$m. The angular divergence is $\Delta \theta_i=1$ mrad and energy spread (FWHM) of $\Delta\varepsilon=42$ MeV. The scattering laser pulse is linearly polarized along $x$ direction, tightly focused at a waist radius of $w_0=3 \mu$m, and Gaussian distributed in temporal dimension at a pulse duration of $\tau=8 T_0$. The laser wavelength is $\lambda_0=1 \mu$m. 

Two-dimension distributions of electron number density with respect to deflection angles of $\theta_x$ and $\theta_y$, corresponding to the intuitive image from the detector of electronic deposition, are shown in the left column of Fig.~\ref{fig2}.  An asymmetry dominated by QSE appears in the electron distribution: in Fig.~\ref{fig2}(a), the electron distribution is oval-shaped, with the major axis 
along the $x$ direction and the minor axis 
along the $y$ direction;
 while it is round-shaped in Figs.~\ref{fig2}(c),~\ref{fig2}(e) and~\ref{fig2}(g) without QSE.
For a more quantitative analysis
we integrate the electron differential angular distributions
in $\theta_{y}$ or $\theta_x$,  and obtain the one-dimension distribution curves of electron density with respect to $\theta_{x}$ or $\theta_y$, respectively, as shown in the right column of Fig.~\ref{fig2}. In Fig.~\ref{fig2}(b), the angle spreads (FWHM) in $x$ and $y$ directions are $\Delta\theta_x=4.28$ mrad and $\Delta\theta_y=1.46$ mrad, respectively, resulting in an asymmetry of $\delta=\Delta\theta_x/\Delta\theta_y\approx3$; in both Fig.~\ref{fig2}(d) and~\ref{fig2}(f), $\Delta\theta_x=\Delta\theta_y=1.46$ mrad; and in Fig.~\ref{fig2}(h), $\Delta\theta_x=\Delta\theta_y=1.12$ mrad. With the initial $\Delta\theta_i=1$ mrad before interaction, the broadening of $\Delta\theta_{x,y}$ (i.e. 0.12 mrad) in Fig.~\ref{fig2}(h) results from the ponderomotive force of ${\bf F}_p=-\bigtriangledown a_0^2/(2\gamma)$ \cite{Quesnel1998}. When RR is included,  the electrons also suffer from radiative energy loss of $\Delta \gamma\sim \alpha a_0 \chi \tau \gamma$ \cite{Ritus1985}, leading to the wider $\Delta\theta_{y}\sim\Delta p_{y}/p_z\propto 1/\gamma$ in Fig.~\ref{fig2}(b) and wider $\Delta\theta_{x,y}$ in Figs.~\ref{fig2}(d) and~\ref{fig2}(f). In this scheme of $\chi\ll1$, $R_c\ll1$ and $a_0\ll\gamma$, the electron deflection angle stemming from radiation loss and pondermotive force is far less than that from QSE. With imaging, such as a Lanex film \cite{Esarey2009}, the asymmetric distribution can be recorded to identify the QSE role.  

\begin{figure}[htb]
 	\includegraphics[width=1\linewidth]{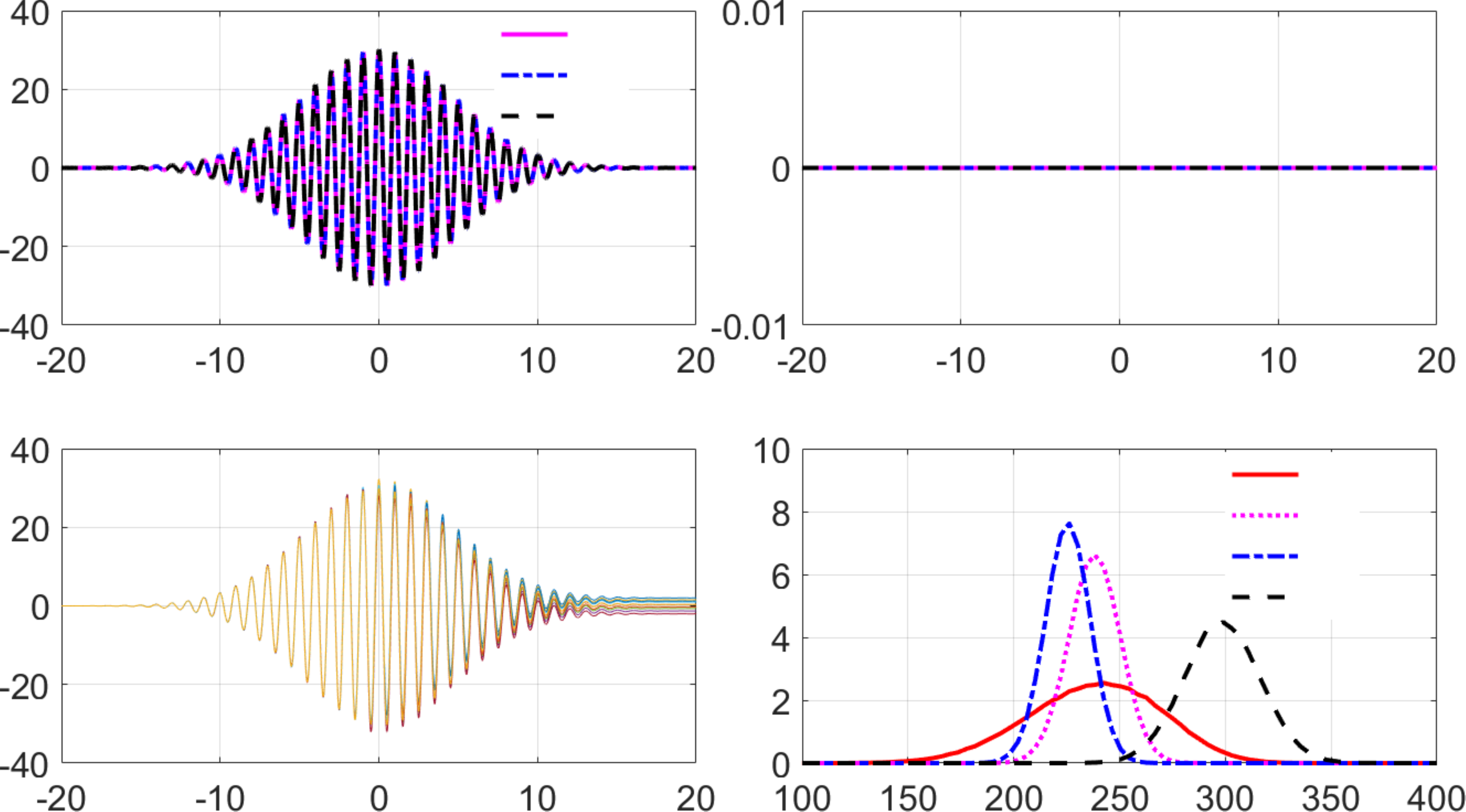}
\begin{picture}(300,0)
       \put(12,138){(a)}
       \put(95,139){\scriptsize MLLM}
       \put(95,132){\scriptsize LLM}
       \put(95,125){\scriptsize w/o RR}
       \put(58,76){\footnotesize $\eta /2\pi$}
       \put(-6,113){\rotatebox{90}{\footnotesize $p_x$}} 
       
       \put(137,138){(b)}
       \put(182,76){\footnotesize $\eta /2\pi$}
       \put(120,113){\rotatebox{90}{\footnotesize $p_y$}} 
       
       \put(12,62){(c)}
       \put(58,2){\footnotesize $\eta /2\pi$}
       \put(-6,43){\rotatebox{90}{\footnotesize $p_x$}}
       
       \put(137,62){(d)}
        \put(135,76){\scriptsize $\times 10^3$}
        \put(221,65){\scriptsize MCM}
        \put(221,57){\scriptsize MLLM}
       \put(221,50){\scriptsize LLM}
       \put(221,43){\scriptsize w/o RR}
        \put(182,2){\footnotesize $\varepsilon$ (MeV)}
       \put(120,33){\rotatebox{90}{\footnotesize ${\rm d} N_e/{\rm d} \varepsilon$}} 
\end{picture}
\caption{The evolution of $p_x$ (a) and $p_y$ (b) with respect to the laser phase $\eta=\omega_0 t -k z$, calculated in MLLM (magenta-solid), LLM (blue-dash-dotted) and without RR effects (black-dashed). (c): The evolution of $p_x$ for 10 sample electrons  in MCM. (d): The final electron energy spectrum in MCM (red-solid), MLLM (magenta-dotted), LLM (blue-dash-dotted) and without RR effects (black-dashed).  } \label{fig3}
\end{figure}
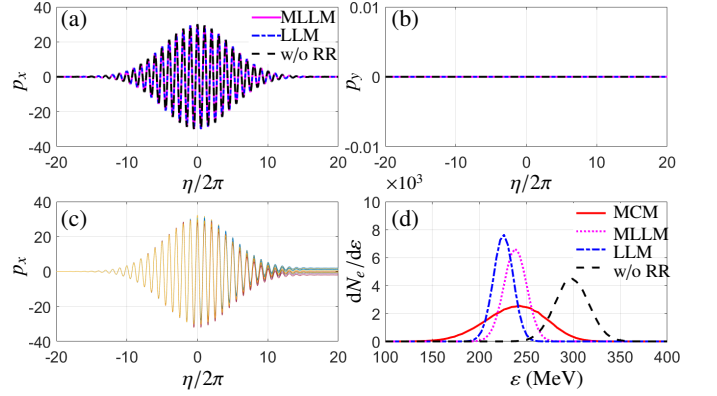

The further explanation of the asymmetric electron distribution from QSE is analyzed in Fig.~\ref{fig3}.  The momenta of a number of sample electrons are calculated.  For simplicity, we set the initial electron momentum along $-z$ direction and the initial position of (0, 0, $c t_0$), with $t_0$ defined as the time when the electron reaches the laser focus. Apart from RR, the increment of electron momentum should be zero, as one can see from the Lorentz equation for electron transverse motion of ${\rm d}{\bf p}_\perp/{\rm d} t=-e({\bf E}_\perp+v\times{\bf B}_\perp)$ with the integral value of ${\bf E}_\perp$ being zero in a normal symmetric laser pulse. As $\gamma\gg 1$, the RR force could be estimated by the leading order of $\gamma^2$ in Eq.~(\ref{LL}) as ${\bf F}_{RR}\approx-2e^4/(3m^2)\gamma^2 {\bf v}[({\bf E}+{\bf v}\times {\bf B})^2-({\bf E}\cdot{\bf v})^2]$. With $|{\bf v}| \approx -v_z \approx 1$, it can be written as ${\bf F}_{RR}\approx-8e^4/(3m^3)\gamma|{\bf E}|^2 {\bf p}$. In this case that ${\bf p}$ evolves according to ${\bf E}$ with a phase delay of $\pi/2$, the integral of the RR force in $x$ direction is nearly zero. Correspondingly, the net transverse momentum increment of an electron passing through the symmetric laser field should also be zero with RR included, in coincident with the numerical result in LLM.  With $g(\chi)\in [0.96,1]$, the electron deflection angle in MLLM is close to that in LLM. Above all, whenever RR is considered or not, in our regime, the final $\Delta p_x$ in Fig.~\ref{fig3}(a) should be zero, and naturally $\Delta p_y$ in Fig.~\ref{fig3}(b) should also be zero due to the linear polarization of the laser pulse. Without QSE, the electron distribution should be symmetric transversely, as shown in Fig.~\ref{fig2}.  

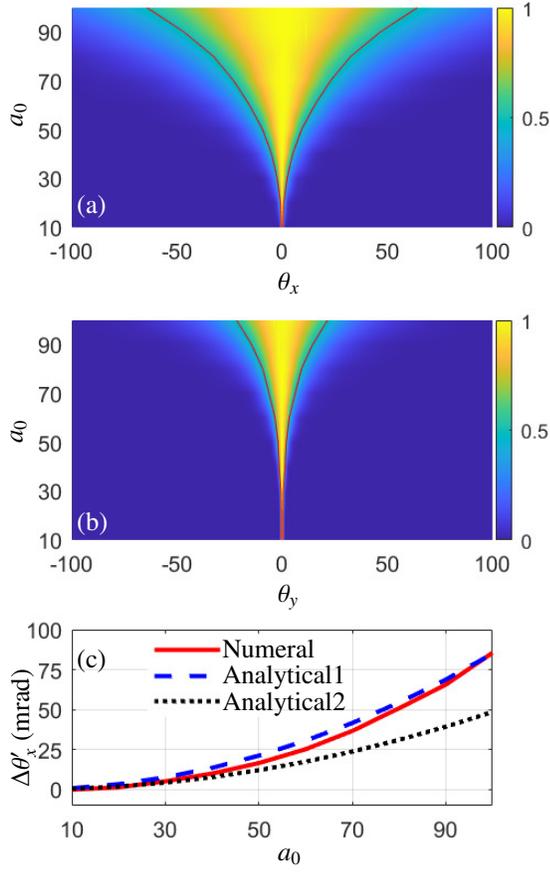
\begin{figure}[t]
 	\includegraphics[width=0.8\linewidth]{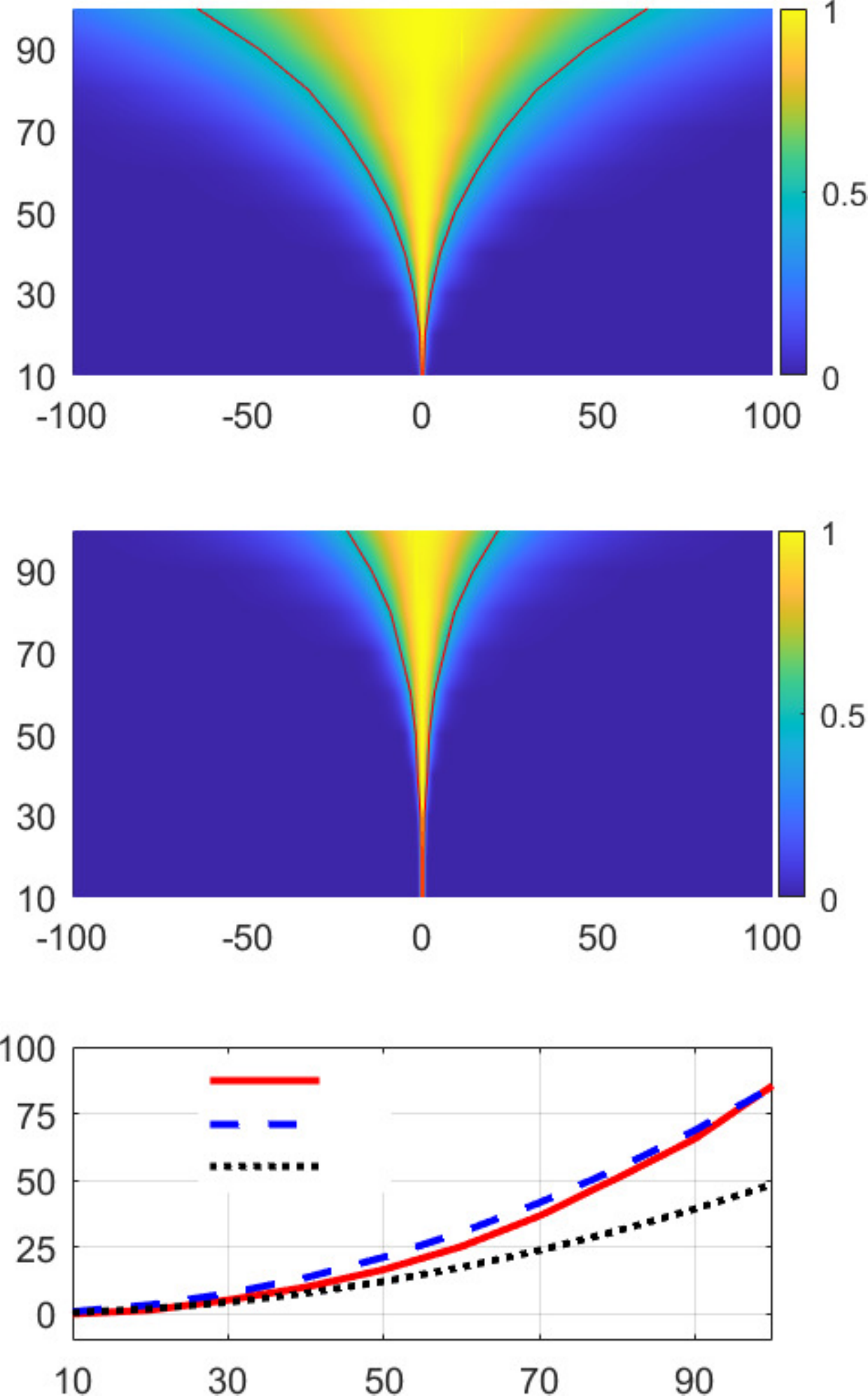}
 \begin{picture}(200,0)
 		\put(20,247){\normalsize {\color{white}(a)}}
	    \put(-5,280){\rotatebox{90}{\normalsize $a_0$}} 
	    \put(96,218){\normalsize $\theta_x$}
	    
	    \put(20,127){\normalsize {\color{white}(b)}}
	    \put(-5,160){\rotatebox{90}{\normalsize $a_0$}} 
	    \put(96,100){\normalsize $\theta_y$}
	    
	    \put(20,75){\normalsize (c)}
	    \put(-5,30){\rotatebox{90}{\normalsize $\Delta \theta_x'$ (mrad)}} 
	    \put(96,2){\normalsize $a_0$}
	    \put(75,79){\normalsize Numeral}
	    \put(75,69){\normalsize Analytical1}
        \put(75,59){\normalsize Analytical2}

\end{picture}
\caption{Distributions of electron density ${\rm d}N_e/{\rm d}\theta_{x} $ vs $\theta_x$ (a) and ${\rm d}N_e/{\rm d}\theta_{y} $ vs $\theta_y$ (b), with $a_0$ increasing from 10 to 100, in MCM. (c) The angle spread caused by QSE obtained numerically with $\Delta \theta_{xN}'=\Delta \theta_x -\Delta \theta_y$ (red-solid) and analytically with $\Delta \theta_{xA1}'=1.75\times2(\omega_0 /m) a_0^2$ (blue-dashed) and $\Delta \theta_{xA2}'=2(\omega_0 /m) a_0^2$ (black-dotted).} \label{fig4}
\end{figure}

The evolutions of $p_x$ of 10 sample electrons in MCM is elaborated in Fig.~\ref{fig3}(c).  Each electron experiences a stochastic radiation process, resulting in a randomly distributed final $p_x$, causing a broadened $\Delta \theta_x$ in Fig.\ref{fig2}(a). 
The broadening effect in one emission could be estimated from $\Delta \theta_x\sim(p_x^i-p_x^\gamma)/p_z=-p_x^\gamma/p_z \lesssim -\chi a_0 /\gamma \approx - 2\omega_0 a_0^2/m\approx4.36$ mrad, where $p_x^i$ and $p_x^\gamma$ are the initial electron momentum component and emitted photon momentum component in the $x$ direction. The number of photons emitted by one electron should be $N_\gamma \sim \alpha a_0 \tau/T_0=1.75$. The overlap of effects from multiple emission lead to a total enlargement of $\Delta \theta_x$ around 3 times larger than $\Delta\theta_y$ in Fig.~\ref{fig2}(a). Therefore, even in quasi-classical regime of $\chi\ll1$ and $R_c\ll1$, we can obtain an obvious angle broadening of $\Delta \theta_x$ dominated by QSE. 

As investigated previously \cite{Arran2019,Neitz2013,Piazza2009,Niel2018,Blackburn2014,Harvey2017}, the final electron energy spectrum can also spread due to the QSE role, see Fig.~\ref{fig3}(d). The spectrum curve of MCM occupies the same mean electron energy of  $240$ MeV with that in MLLM, but with a wider spread of $\Delta\varepsilon=75$ MeV than that of $\Delta\varepsilon=28$ MeV in MLLM. In LLM, a similar energy spread $\Delta\varepsilon=24$ MeV is obtained, but the mean electron energy is lower ($226$ MeV) due to the overestimation of radiation loss. The result that the energy spread is reduced by the RR classically, whereas enlarged by QSE significantly, is consistent with the conclusion of Ref.~\cite{Neitz2013}. The larger energy loss in LLM, cannot lead to a distinguishable difference in electron distribution from MLLM owing to the equivalent overestimation of the transverse momentum loss, as shown in Eq.~(\ref{MLL}). Comparatively, the asymmetric electron angular distribution can be measured more easily than the electron energy spread, particularly considering the fluctuation and statistical uncertainty in experiments.

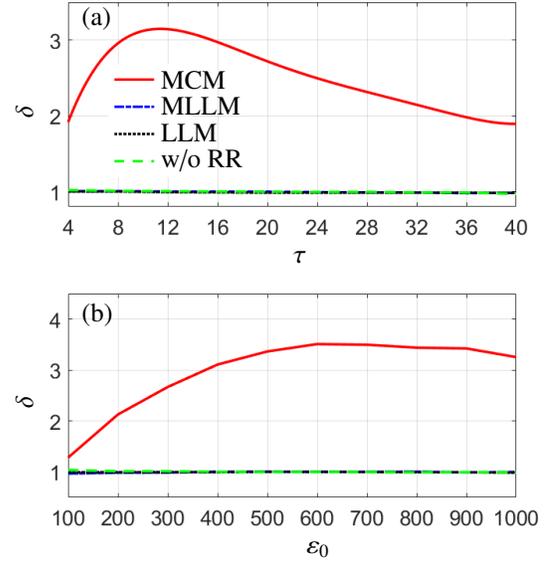
\begin{figure}[t]
 	\includegraphics[width=0.75\linewidth]{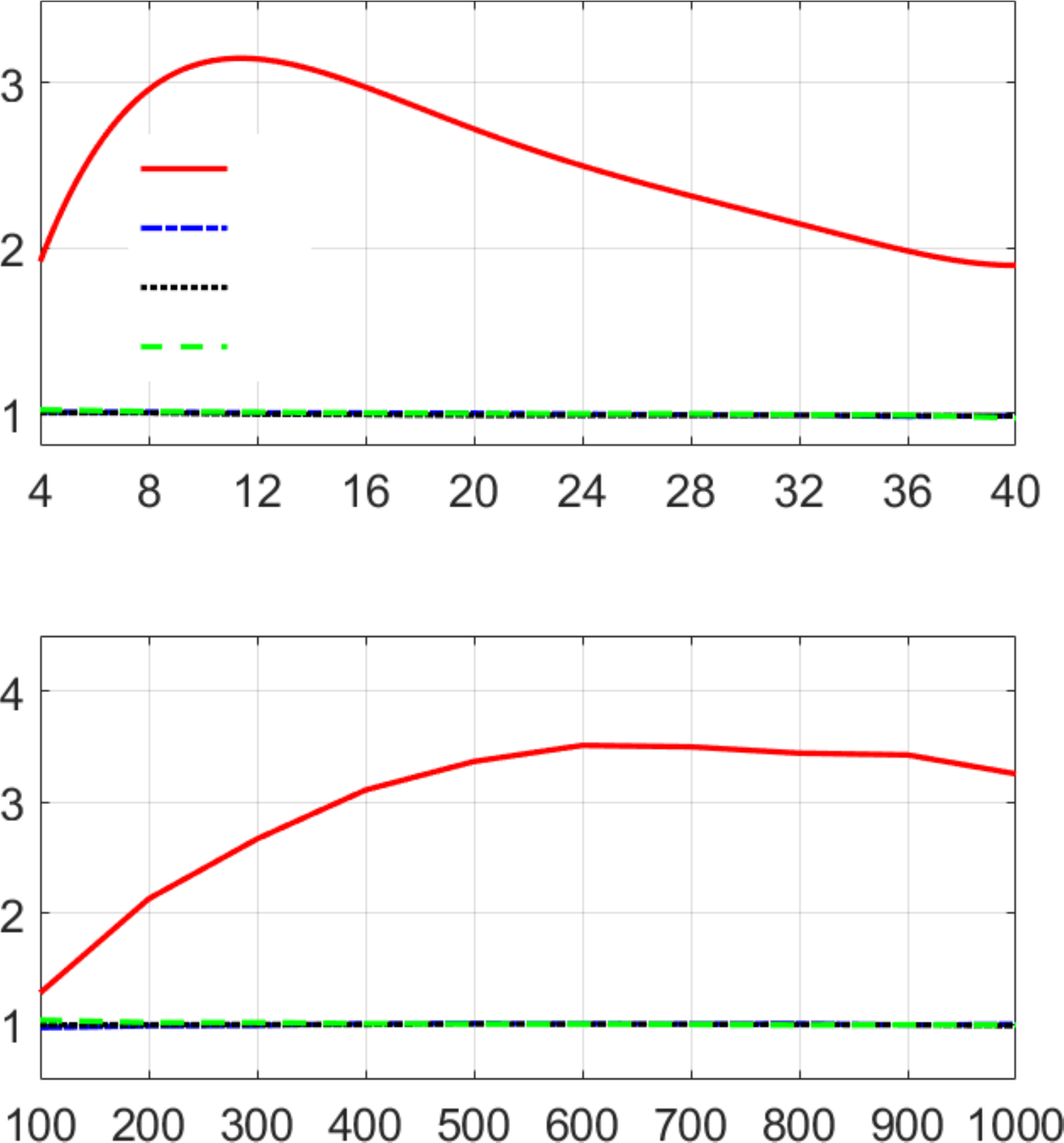}
 \begin{picture}(200,0)
 		\put(20,200){\normalsize {(a)}}
	    \put(-5,165){\rotatebox{90}{\normalsize $\delta$}} 
	    \put(100,110){\normalsize $\tau$}
	    \put(50,176){\normalsize MCM}
	    \put(50,166){\normalsize MLLM}
	    \put(50,156){\normalsize LLM}
	    \put(50,146){\normalsize w/o RR}
	    
	    \put(20,88){\normalsize {(b)}}
	    \put(-5,55){\rotatebox{90}{\normalsize $\delta$}} 
	    \put(105,0){\normalsize $\varepsilon_0$}

\end{picture}
\caption{Impact of laser pulse duration $\tau$ (a) and initial electron kinetic energy $\varepsilon_0$ (b) on the ration of $\delta=\Delta \theta_x/\Delta \theta_y$. } \label{fig5}
\end{figure}

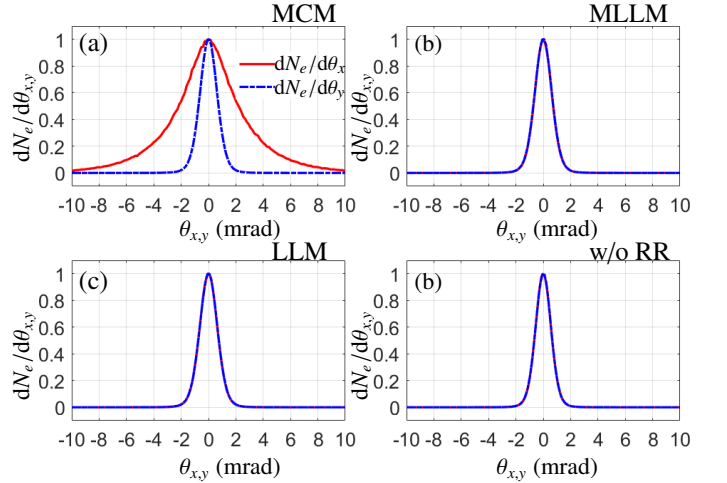
\begin{figure}[htb]
 	\includegraphics[width=1\linewidth]{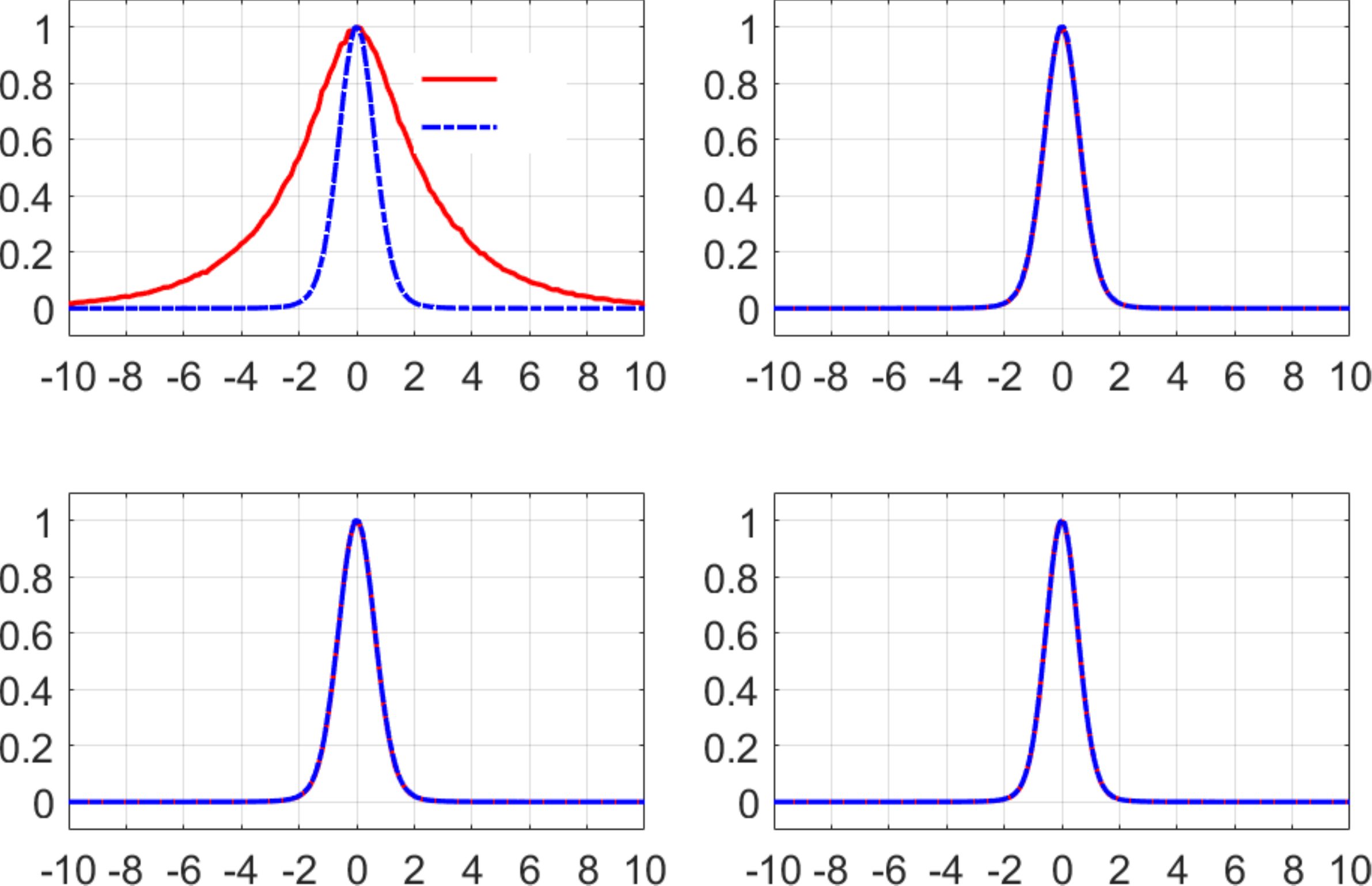}
 \begin{picture}(200,0)
 		\put(-8,160){\normalsize {(a)}}
 		\put(66,153){\footnotesize ${\rm d}N_e/{\rm d}\theta_{x}$} 
       \put(66,144){\footnotesize ${\rm d}N_e/{\rm d}\theta_{y}$}
	    \put(-33,118){\rotatebox{90}{\small ${\rm d}N_e/{\rm d}\theta_{x,y}$}} 
	    \put(30,90){\small $\theta_{x,y}$ (mrad)}
	    \put(65,171){\normalsize MCM}
	    
	    
	    \put(119,160){(b)}
	   \put(95,118){\rotatebox{90}{\small ${\rm d}N_e/{\rm d}\theta_{x,y}$}} 
	    \put(150,90){\small $\theta_{x,y}$ (mrad)}
	    \put(185,171){\normalsize MLLM}
	    
	    \put(-8,70){\normalsize {(c)}}
	    \put(-33,28){\rotatebox{90}{\small ${\rm d}N_e/{\rm d}\theta_{x,y}$}} 
	    \put(30,0){\small $\theta_{x,y}$ (mrad)}
	    \put(65,81){\normalsize LLM}
	    
	    
	    \put(119,70){(b)}
	   \put(95,28){\rotatebox{90}{\small ${\rm d}N_e/{\rm d}\theta_{x,y}$}} 
	    \put(150,0){\small $\theta_{x,y}$ (mrad)}
	    \put(185,81){\normalsize w/o RR}

\end{picture}
\caption{Integrated one-dimension distribution of electron density of ${\rm d}N_e/{\rm d}\theta_{x,y} $ (mrad$^{-1}$) vs $\theta_x$ (red-solid) or $\theta_y$ (blue-dash-dotted) calculated in MCM (a), MLLM (b) and LLM (c), respectively; and without RR (d). The initial electron energy spread is $\sigma_\varepsilon=0.5 \varepsilon_0=150$ MeV, and other laser and electron beam parameters are the same as in Fig.~\ref{fig2}.} \label{fig6}
\end{figure}

The influences of laser and electron beam parameters on the QSE signature are discussed below to examine its robustness and to clarify the requirements for  experimental observation. In Fig.~\ref{fig4}, increasing of $a_0$ from 10 to 100, $\Delta \theta_y$ grows from 1 mrad to 43.1 mrad, and $\Delta \theta_x$ from 1 mrad to 128.6 mrad in a faster pattern. The increment of $\Delta \theta_y$ indicates the tendency of the growing transverse pondermotive force as the corresponding deflection angle of $\theta_p\sim |{\bf F}_{p\perp}|\tau/\gamma\propto(a_0^2/\gamma^2)(\tau/w_0)$ \cite{Quesnel1998,Liyf2018}. The $\Delta \theta_x$ results from a combination of pondermotive force effect and QSE, sensitively dependent on $a_0$ under the domination of the latter. The calculation about the broadening of deflection angle by QSE is performed, in Fig.~\ref{fig4}(c), numerally and analytically. While analytical results taking into account one or 1.75 emission times are calculated, the latter matches well with the numerical one even with an increasing number of emission $N_\gamma\propto a_0$, which can be explained in Fig.~\ref{fig5}. It should be noted that the requirement  crucial to observe the QSE signature in this quasi-classical regime reads 
\begin{eqnarray}\label{theta}
\Delta \theta_i \lesssim 2(\omega_0/m) a_0^2\sim 10^{-6}a_0^2.
\end{eqnarray}
Otherwise, the $\Delta \theta_x'$ would be submerged by the initial angular divergence $\Delta \theta_i$. This requirement gives a minimum of  $a_0\approx20$ (or $I_0\approx5.5\times 10^{20}$W/cm$^2$) for currently available angular divergence of the electron beam \cite{Gonsalves2019}.

The impacts of laser pulse duration and initial electron mean energy are illustrated in Figs.~\ref{fig5}(a) and \ref{fig5}(b). With the duration $\tau$ changing from 4 $T_0$ to 40 $T_0$, the asymmetry $\delta$ rises first from 1.96 to 3.15 at $\tau=11.5 T_0$, and then declines to 1.89 at $\tau=40 T_0$. It is also demonstrated from the curve in Fig.~\ref{fig5}(a) that the deflection effect from QSE (corresponding to $\Delta \theta_x'$) is strengthened at a decreasing speed with the growth of radiation number, which indicates a reasonable estimation of $\Delta\theta_x'\approx1.75\times 2 (\omega_0/m) a_0^2$ as shown in Fig.~\ref{fig4}(c). While the deflection effect resulting from transverse pondermotive force (corresponding to $\Delta \theta_y$) is $\theta_p\propto\tau\propto N_\gamma$  \cite{Quesnel1998}, the asymmetry tends to be counteracted and weakened with abundant photon emissions. To get an apparent asymmetry $\delta$, a moderate initial electron energy $\varepsilon_0$ is necessary, even though $\Delta \theta_{x}'\sim2(\omega_0/m) a_0^2$ is independent of $\varepsilon_0$, see Fig.~\ref{fig5}(b). On the one hand, $\varepsilon_0$ should be large enough to make $\chi > 0.01$ for a photon spectrum wide enough, which is necessary for QSE. On the other, it should also be not too large to ensure $\chi \ll 1$ to avoid momentous radiative loss.

For experimental feasibility, we also consider a case with a larger initial electron energy spread (150 MeV) and show the results in Fig.~\ref{fig6}. The asymmetry of the electron distribution is stable with respect to that in Fig.~\ref{fig2}, since the condition of $\Delta \theta_i \lesssim 2(\omega_0/m) a_0^2$ is fulfilled. 

In conclusion, we have investigated the QSE effects of photon emission on the dynamics of an electron beam head-on colliding with a linearly polarized laser pulse in a quasi-classical regime of $\chi\ll1$ and $R_c\ll1$. Under the condition of $\Delta \theta_i\lesssim 10^{-6}a_0^2$, even when the radiation loss is far less than the electron kinetic energy, the QSE could be elicited and distinguished by the asymmetry of the final electron angular distribution between the laser polarization direction and the other orthogonal direction. This QSE signature could be observed intuitively on the image from detector of electron deposition, only by a single-shot. It provides a feasible scheme to test one of the fundamental quantum properties, the stochasticity nature of photon emission, with laser intensity $I_0\lesssim 10^{21}$ and electron energy of hundreds of MeV currently available in experiments . 
 
{\it Acknowledgement:} 
 This work is supported by the National Natural Science Foundation of China (Grants Nos. 11804269, U1830128, 11875214,  11775302, and 11975182), the National Key R\&D Program of China (Grant No. 2018YFA0404801), NSAF Joint Fund (Grant No.
U1830128), the Strategic Priority Research Program of Chinese Academy of Sciences (Grant Nos. XDA25050300, XDA16010200), and the Fundamental Research Funds for the Central Universities, and the Research Funds of Renmin University of China (Grant No. 20XNLG01).
\bibliography{reference.bib}

\begin{thebibliography}{47}%
\makeatletter
\providecommand \@ifxundefined [1]{%
 \@ifx{#1\undefined}
}%
\providecommand \@ifnum [1]{%
 \ifnum #1\expandafter \@firstoftwo
 \else \expandafter \@secondoftwo
 \fi
}%
\providecommand \@ifx [1]{%
 \ifx #1\expandafter \@firstoftwo
 \else \expandafter \@secondoftwo
 \fi
}%
\providecommand \natexlab [1]{#1}%
\providecommand \enquote  [1]{``#1''}%
\providecommand \bibnamefont  [1]{#1}%
\providecommand \bibfnamefont [1]{#1}%
\providecommand \citenamefont [1]{#1}%
\providecommand \href@noop [0]{\@secondoftwo}%
\providecommand \href [0]{\begingroup \@sanitize@url \@href}%
\providecommand \@href[1]{\@@startlink{#1}\@@href}%
\providecommand \@@href[1]{\endgroup#1\@@endlink}%
\providecommand \@sanitize@url [0]{\catcode `\\12\catcode `\$12\catcode
  `\&12\catcode `\#12\catcode `\^12\catcode `\_12\catcode `\%12\relax}%
\providecommand \@@startlink[1]{}%
\providecommand \@@endlink[0]{}%
\providecommand \url  [0]{\begingroup\@sanitize@url \@url }%
\providecommand \@url [1]{\endgroup\@href {#1}{\urlprefix }}%
\providecommand \urlprefix  [0]{URL }%
\providecommand \Eprint [0]{\href }%
\providecommand \doibase [0]{http://dx.doi.org/}%
\providecommand \selectlanguage [0]{\@gobble}%
\providecommand \bibinfo  [0]{\@secondoftwo}%
\providecommand \bibfield  [0]{\@secondoftwo}%
\providecommand \translation [1]{[#1]}%
\providecommand \BibitemOpen [0]{}%
\providecommand \bibitemStop [0]{}%
\providecommand \bibitemNoStop [0]{.\EOS\space}%
\providecommand \EOS [0]{\spacefactor3000\relax}%
\providecommand \BibitemShut  [1]{\csname bibitem#1\endcsname}%
\let\auto@bib@innerbib\@empty
\bibitem [{\citenamefont {Jackson}(1998)}]{Jackson1998}%
  \BibitemOpen
  \bibfield  {author} {\bibinfo {author} {\bibfnamefont {John~David}\
  \bibnamefont {Jackson}},\ }\href@noop {} {\emph {\bibinfo {title} {Classical
  Electrondynamics}}},\ \bibinfo {edition} {3rd}\ ed.\ (\bibinfo  {publisher}
  {Wiley},\ \bibinfo {year} {1998})\ Chap.~\bibinfo {chapter} {16}\BibitemShut
  {NoStop}%
\bibitem [{\citenamefont {Berestetskii}\ \emph {et~al.}(1982)\citenamefont
  {Berestetskii}, \citenamefont {Lifshitz},\ and\ \citenamefont
  {Pitaevskii}}]{Berestetskii1982}%
  \BibitemOpen
  \bibfield  {author} {\bibinfo {author} {\bibfnamefont {V.~B.}\ \bibnamefont
  {Berestetskii}}, \bibinfo {author} {\bibfnamefont {E.~M.}\ \bibnamefont
  {Lifshitz}}, \ and\ \bibinfo {author} {\bibfnamefont {L.~P.}\ \bibnamefont
  {Pitaevskii}},\ }\href@noop {} {\emph {\bibinfo {title} {Quantum
  Electrodynamics}}}\ (\bibinfo  {publisher} {Pergamon},\ \bibinfo {address}
  {Oxford},\ \bibinfo {year} {1982})\BibitemShut {NoStop}%
\bibitem [{\citenamefont {Abraham}()}]{Abraham1905}%
  \BibitemOpen
  \bibfield  {author} {\bibinfo {author} {\bibfnamefont {M.}~\bibnamefont
  {Abraham}},\ }\href@noop {} {\emph {\bibinfo {title} {Theorie der
  Elektrizitat, Vol. II: Elektromagnetische Theorie der Strahlung}}}\ (\bibinfo
   {publisher} {Teubner},\ \bibinfo {address} {Leipzig})\BibitemShut {NoStop}%
\bibitem [{\citenamefont {Lorentz}(1909)}]{Lorentz1909}%
  \BibitemOpen
  \bibfield  {author} {\bibinfo {author} {\bibfnamefont {A.}~\bibnamefont
  {Lorentz}},\ }\href@noop {} {\emph {\bibinfo {title} {The Theory of
  Electrons}}}\ (\bibinfo  {publisher} {Teubner},\ \bibinfo {address}
  {Leipzig},\ \bibinfo {year} {1909})\BibitemShut {NoStop}%
\bibitem [{\citenamefont {Dirac}(1938)}]{Dirac1938}%
  \BibitemOpen
  \bibfield  {author} {\bibinfo {author} {\bibfnamefont {P.~A.~M.}\
  \bibnamefont {Dirac}},\ }\bibfield  {title} {\enquote {\bibinfo {title}
  {Classical theory of radiative electrons},}\ }\href@noop {} {\bibfield
  {journal} {\bibinfo  {journal} {Proc. R. Soc. London Ser. A}\ }\textbf
  {\bibinfo {volume} {167}} (\bibinfo {year} {1938})}\BibitemShut {NoStop}%
\bibitem [{\citenamefont {Landau}\ and\ \citenamefont
  {Lifshitz}(1975)}]{Landau1975}%
  \BibitemOpen
  \bibfield  {author} {\bibinfo {author} {\bibfnamefont {L~D}\ \bibnamefont
  {Landau}}\ and\ \bibinfo {author} {\bibfnamefont {E~M}\ \bibnamefont
  {Lifshitz}},\ }\href@noop {} {\emph {\bibinfo {title} {The Classical Theory
  of Fields}}}\ (\bibinfo  {publisher} {Elsevier, Oxford},\ \bibinfo {year}
  {1975})\BibitemShut {NoStop}%
\bibitem [{\citenamefont {Yoffe}\ \emph {et~al.}(2015)\citenamefont {Yoffe},
  \citenamefont {Kravets}, \citenamefont {Noble},\ and\ \citenamefont
  {Jaroszynski}}]{Yoffe2015}%
  \BibitemOpen
  \bibfield  {author} {\bibinfo {author} {\bibfnamefont {Samuel~R}\
  \bibnamefont {Yoffe}}, \bibinfo {author} {\bibfnamefont {Yevgen}\
  \bibnamefont {Kravets}}, \bibinfo {author} {\bibfnamefont {Adam}\
  \bibnamefont {Noble}}, \ and\ \bibinfo {author} {\bibfnamefont {Dino~A}\
  \bibnamefont {Jaroszynski}},\ }\bibfield  {title} {\enquote {\bibinfo {title}
  {Longitudinal and transverse cooling of relativistic electron beams in
  intense laser pulses},}\ }\href {\doibase 10.1088/1367-2630/17/5/053025}
  {\bibfield  {journal} {\bibinfo  {journal} {New J. Phys.}\ }\textbf {\bibinfo
  {volume} {17}},\ \bibinfo {pages} {053025} (\bibinfo {year}
  {2015})}\BibitemShut {NoStop}%
\bibitem [{\citenamefont {Sokolov}\ \emph {et~al.}(2010)\citenamefont
  {Sokolov}, \citenamefont {Nees}, \citenamefont {Yanovsky}, \citenamefont
  {Naumova},\ and\ \citenamefont {Mourou}}]{Sokolov2010}%
  \BibitemOpen
  \bibfield  {author} {\bibinfo {author} {\bibfnamefont {Igor~V.}\ \bibnamefont
  {Sokolov}}, \bibinfo {author} {\bibfnamefont {John~A.}\ \bibnamefont {Nees}},
  \bibinfo {author} {\bibfnamefont {Victor~P.}\ \bibnamefont {Yanovsky}},
  \bibinfo {author} {\bibfnamefont {Natalia~M.}\ \bibnamefont {Naumova}}, \
  and\ \bibinfo {author} {\bibfnamefont {G\'erard~A.}\ \bibnamefont {Mourou}},\
  }\bibfield  {title} {\enquote {\bibinfo {title} {Emission and its
  back-reaction accompanying electron motion in relativistically strong and
  qed-strong pulsed laser fields},}\ }\href {\doibase
  10.1103/PhysRevE.81.036412} {\bibfield  {journal} {\bibinfo  {journal} {Phys.
  Rev. E}\ }\textbf {\bibinfo {volume} {81}},\ \bibinfo {pages} {036412}
  (\bibinfo {year} {2010})}\BibitemShut {NoStop}%
\bibitem [{\citenamefont {Erber}(1966)}]{Erber1966}%
  \BibitemOpen
  \bibfield  {author} {\bibinfo {author} {\bibfnamefont {Thomas}\ \bibnamefont
  {Erber}},\ }\bibfield  {title} {\enquote {\bibinfo {title} {High-energy
  electromagnetic conversion processes in intense magnetic fields},}\ }\href
  {\doibase 10.1103/RevModPhys.38.626} {\bibfield  {journal} {\bibinfo
  {journal} {Rev. Mod. Phys.}\ }\textbf {\bibinfo {volume} {38}},\ \bibinfo
  {pages} {626--659} (\bibinfo {year} {1966})}\BibitemShut {NoStop}%
\bibitem [{\citenamefont {Neitz}\ and\ \citenamefont
  {Di~Piazza}(2013)}]{Neitz2013}%
  \BibitemOpen
  \bibfield  {author} {\bibinfo {author} {\bibfnamefont {N.}~\bibnamefont
  {Neitz}}\ and\ \bibinfo {author} {\bibfnamefont {A.}~\bibnamefont
  {Di~Piazza}},\ }\bibfield  {title} {\enquote {\bibinfo {title} {Stochasticity
  effects in quantum radiation reaction},}\ }\href {\doibase
  10.1103/PhysRevLett.111.054802} {\bibfield  {journal} {\bibinfo  {journal}
  {Phys. Rev. Lett.}\ }\textbf {\bibinfo {volume} {111}},\ \bibinfo {pages}
  {054802} (\bibinfo {year} {2013})}\BibitemShut {NoStop}%
\bibitem [{\citenamefont {Neitz}\ and\ \citenamefont
  {Di~Piazza}(2014)}]{Neitz2014}%
  \BibitemOpen
  \bibfield  {author} {\bibinfo {author} {\bibfnamefont {N.}~\bibnamefont
  {Neitz}}\ and\ \bibinfo {author} {\bibfnamefont {A.}~\bibnamefont
  {Di~Piazza}},\ }\bibfield  {title} {\enquote {\bibinfo {title} {Electron-beam
  dynamics in a strong laser field including quantum radiation reaction},}\
  }\href {\doibase 10.1103/PhysRevA.90.022102} {\bibfield  {journal} {\bibinfo
  {journal} {Phys. Rev. A}\ }\textbf {\bibinfo {volume} {90}},\ \bibinfo
  {pages} {022102} (\bibinfo {year} {2014})}\BibitemShut {NoStop}%
\bibitem [{\citenamefont {Wan}\ \emph {et~al.}(2019)\citenamefont {Wan},
  \citenamefont {Xue}, \citenamefont {Dou}, \citenamefont {Hatsagortsyan},
  \citenamefont {Yan}, \citenamefont {Khikhlukha}, \citenamefont {Bulanov},
  \citenamefont {Korn}, \citenamefont {Zhao}, \citenamefont {Xu},\ and\
  \citenamefont {Li}}]{Wan2019}%
  \BibitemOpen
  \bibfield  {author} {\bibinfo {author} {\bibfnamefont {Feng}\ \bibnamefont
  {Wan}}, \bibinfo {author} {\bibfnamefont {Kun}\ \bibnamefont {Xue}}, \bibinfo
  {author} {\bibfnamefont {Zhen-Ke}\ \bibnamefont {Dou}}, \bibinfo {author}
  {\bibfnamefont {Karen~Z}\ \bibnamefont {Hatsagortsyan}}, \bibinfo {author}
  {\bibfnamefont {Wenchao}\ \bibnamefont {Yan}}, \bibinfo {author}
  {\bibfnamefont {Danila}\ \bibnamefont {Khikhlukha}}, \bibinfo {author}
  {\bibfnamefont {Sergei~V}\ \bibnamefont {Bulanov}}, \bibinfo {author}
  {\bibfnamefont {Georg}\ \bibnamefont {Korn}}, \bibinfo {author}
  {\bibfnamefont {Yong-Tao}\ \bibnamefont {Zhao}}, \bibinfo {author}
  {\bibfnamefont {Zhong-Feng}\ \bibnamefont {Xu}}, \ and\ \bibinfo {author}
  {\bibfnamefont {Jian-Xing}\ \bibnamefont {Li}},\ }\bibfield  {title}
  {\enquote {\bibinfo {title} {Imprint of the stochastic nature of photon
  emission by electrons on the proton energy spectra in the laser-plasma
  interaction},}\ }\href {\doibase 10.1088/1361-6587/ab2b2c} {\bibfield
  {journal} {\bibinfo  {journal} {Plasma Phys. Contr. F.}\ }\textbf {\bibinfo
  {volume} {61}},\ \bibinfo {pages} {084010} (\bibinfo {year}
  {2019})}\BibitemShut {NoStop}%
\bibitem [{\citenamefont {Blackburn}\ \emph {et~al.}(2014)\citenamefont
  {Blackburn}, \citenamefont {Ridgers}, \citenamefont {Kirk},\ and\
  \citenamefont {Bell}}]{Blackburn2014}%
  \BibitemOpen
  \bibfield  {author} {\bibinfo {author} {\bibfnamefont {T.~G.}\ \bibnamefont
  {Blackburn}}, \bibinfo {author} {\bibfnamefont {C.~P.}\ \bibnamefont
  {Ridgers}}, \bibinfo {author} {\bibfnamefont {J.~G.}\ \bibnamefont {Kirk}}, \
  and\ \bibinfo {author} {\bibfnamefont {A.~R.}\ \bibnamefont {Bell}},\
  }\bibfield  {title} {\enquote {\bibinfo {title} {Quantum radiation reaction
  in laser--electron-beam collisions},}\ }\href {\doibase
  10.1103/PhysRevLett.112.015001} {\bibfield  {journal} {\bibinfo  {journal}
  {Phys. Rev. Lett.}\ }\textbf {\bibinfo {volume} {112}},\ \bibinfo {pages}
  {015001} (\bibinfo {year} {2014})}\BibitemShut {NoStop}%
\bibitem [{\citenamefont {Harvey}\ \emph {et~al.}(2017)\citenamefont {Harvey},
  \citenamefont {Gonoskov}, \citenamefont {Ilderton},\ and\ \citenamefont
  {Marklund}}]{Harvey2017}%
  \BibitemOpen
  \bibfield  {author} {\bibinfo {author} {\bibfnamefont {C.~N.}\ \bibnamefont
  {Harvey}}, \bibinfo {author} {\bibfnamefont {A.}~\bibnamefont {Gonoskov}},
  \bibinfo {author} {\bibfnamefont {A.}~\bibnamefont {Ilderton}}, \ and\
  \bibinfo {author} {\bibfnamefont {M.}~\bibnamefont {Marklund}},\ }\bibfield
  {title} {\enquote {\bibinfo {title} {Quantum quenching of radiation losses in
  short laser pulses},}\ }\href {\doibase 10.1103/PhysRevLett.118.105004}
  {\bibfield  {journal} {\bibinfo  {journal} {Phys. Rev. Lett.}\ }\textbf
  {\bibinfo {volume} {118}},\ \bibinfo {pages} {105004} (\bibinfo {year}
  {2017})}\BibitemShut {NoStop}%
\bibitem [{\citenamefont {Di~Piazza}\ \emph {et~al.}(2009)\citenamefont
  {Di~Piazza}, \citenamefont {Hatsagortsyan},\ and\ \citenamefont
  {Keitel}}]{Piazza2009}%
  \BibitemOpen
  \bibfield  {author} {\bibinfo {author} {\bibfnamefont {A.}~\bibnamefont
  {Di~Piazza}}, \bibinfo {author} {\bibfnamefont {K.~Z.}\ \bibnamefont
  {Hatsagortsyan}}, \ and\ \bibinfo {author} {\bibfnamefont {C.~H.}\
  \bibnamefont {Keitel}},\ }\bibfield  {title} {\enquote {\bibinfo {title}
  {Strong signatures of radiation reaction below the radiation-dominated
  regime},}\ }\href {\doibase 10.1103/PhysRevLett.102.254802} {\bibfield
  {journal} {\bibinfo  {journal} {Phys. Rev. Lett.}\ }\textbf {\bibinfo
  {volume} {102}},\ \bibinfo {pages} {254802} (\bibinfo {year}
  {2009})}\BibitemShut {NoStop}%
\bibitem [{\citenamefont {Niel}\ \emph {et~al.}(2018)\citenamefont {Niel},
  \citenamefont {Riconda}, \citenamefont {Amiranoff}, \citenamefont {Duclous},\
  and\ \citenamefont {Grech}}]{Niel2018}%
  \BibitemOpen
  \bibfield  {author} {\bibinfo {author} {\bibfnamefont {F.}~\bibnamefont
  {Niel}}, \bibinfo {author} {\bibfnamefont {C.}~\bibnamefont {Riconda}},
  \bibinfo {author} {\bibfnamefont {F.}~\bibnamefont {Amiranoff}}, \bibinfo
  {author} {\bibfnamefont {R.}~\bibnamefont {Duclous}}, \ and\ \bibinfo
  {author} {\bibfnamefont {M.}~\bibnamefont {Grech}},\ }\bibfield  {title}
  {\enquote {\bibinfo {title} {From quantum to classical modeling of radiation
  reaction: A focus on stochasticity effects},}\ }\href {\doibase
  10.1103/PhysRevE.97.043209} {\bibfield  {journal} {\bibinfo  {journal} {Phys.
  Rev. E}\ }\textbf {\bibinfo {volume} {97}},\ \bibinfo {pages} {043209}
  (\bibinfo {year} {2018})}\BibitemShut {NoStop}%
\bibitem [{\citenamefont {Li}\ \emph {et~al.}(2017)\citenamefont {Li},
  \citenamefont {Chen}, \citenamefont {Hatsagortsyan},\ and\ \citenamefont
  {Keitel}}]{Li2017}%
  \BibitemOpen
  \bibfield  {author} {\bibinfo {author} {\bibfnamefont {Jian-Xing}\
  \bibnamefont {Li}}, \bibinfo {author} {\bibfnamefont {Yue-Yue}\ \bibnamefont
  {Chen}}, \bibinfo {author} {\bibfnamefont {Karen~Z.}\ \bibnamefont
  {Hatsagortsyan}}, \ and\ \bibinfo {author} {\bibfnamefont {Christoph~H.}\
  \bibnamefont {Keitel}},\ }\bibfield  {title} {\enquote {\bibinfo {title}
  {Angle-resolved stochastic photon emission in the quantum radiation-dominated
  regime},}\ }\href@noop {} {\bibfield  {journal} {\bibinfo  {journal} {Sci.
  Rep.}\ }\textbf {\bibinfo {volume} {7}} (\bibinfo {year} {2017})}\BibitemShut
  {NoStop}%
\bibitem [{\citenamefont {Li}\ \emph {et~al.}(2018)\citenamefont {Li},
  \citenamefont {Zhao}, \citenamefont {Hatsagortsyan}, \citenamefont {Keitel},\
  and\ \citenamefont {Li}}]{Liyf2018}%
  \BibitemOpen
  \bibfield  {author} {\bibinfo {author} {\bibfnamefont {Yan-Fei}\ \bibnamefont
  {Li}}, \bibinfo {author} {\bibfnamefont {Yong-Tao}\ \bibnamefont {Zhao}},
  \bibinfo {author} {\bibfnamefont {Karen~Z.}\ \bibnamefont {Hatsagortsyan}},
  \bibinfo {author} {\bibfnamefont {Christoph~H.}\ \bibnamefont {Keitel}}, \
  and\ \bibinfo {author} {\bibfnamefont {Jian-Xing}\ \bibnamefont {Li}},\
  }\bibfield  {title} {\enquote {\bibinfo {title}
  {Electron-angular-distribution reshaping in the quantum radiation-dominated
  regime},}\ }\href {\doibase 10.1103/PhysRevA.98.052120} {\bibfield  {journal}
  {\bibinfo  {journal} {Phys. Rev. A}\ }\textbf {\bibinfo {volume} {98}},\
  \bibinfo {pages} {052120} (\bibinfo {year} {2018})}\BibitemShut {NoStop}%
\bibitem [{\citenamefont {Baird}\ \emph {et~al.}(2019)\citenamefont {Baird},
  \citenamefont {Murphy}, \citenamefont {Blackburn}, \citenamefont {Ilderton},
  \citenamefont {Mangles}, \citenamefont {Marklund},\ and\ \citenamefont
  {Ridgers}}]{Baird2019}%
  \BibitemOpen
  \bibfield  {author} {\bibinfo {author} {\bibfnamefont {C~D}\ \bibnamefont
  {Baird}}, \bibinfo {author} {\bibfnamefont {C~D}\ \bibnamefont {Murphy}},
  \bibinfo {author} {\bibfnamefont {T~G}\ \bibnamefont {Blackburn}}, \bibinfo
  {author} {\bibfnamefont {A}~\bibnamefont {Ilderton}}, \bibinfo {author}
  {\bibfnamefont {S~P~D}\ \bibnamefont {Mangles}}, \bibinfo {author}
  {\bibfnamefont {M}~\bibnamefont {Marklund}}, \ and\ \bibinfo {author}
  {\bibfnamefont {C~P}\ \bibnamefont {Ridgers}},\ }\bibfield  {title} {\enquote
  {\bibinfo {title} {Realising single-shot measurements of quantum radiation
  reaction in high-intensity lasers},}\ }\href {\doibase
  10.1088/1367-2630/ab1baf} {\bibfield  {journal} {\bibinfo  {journal} {New J.
  Phys.}\ }\textbf {\bibinfo {volume} {21}},\ \bibinfo {pages} {053030}
  (\bibinfo {year} {2019})}\BibitemShut {NoStop}%
\bibitem [{\citenamefont {Bashinov}\ \emph {et~al.}(2015)\citenamefont
  {Bashinov}, \citenamefont {Kim},\ and\ \citenamefont
  {Sergeev}}]{Bashinov2015}%
  \BibitemOpen
  \bibfield  {author} {\bibinfo {author} {\bibfnamefont {A.~V.}\ \bibnamefont
  {Bashinov}}, \bibinfo {author} {\bibfnamefont {A.~V.}\ \bibnamefont {Kim}}, \
  and\ \bibinfo {author} {\bibfnamefont {A.~M.}\ \bibnamefont {Sergeev}},\
  }\bibfield  {title} {\enquote {\bibinfo {title} {Impact of quantum effects on
  relativistic electron motion in a chaotic regime},}\ }\href {\doibase
  10.1103/PhysRevE.92.043105} {\bibfield  {journal} {\bibinfo  {journal} {Phys.
  Rev. E}\ }\textbf {\bibinfo {volume} {92}},\ \bibinfo {pages} {043105}
  (\bibinfo {year} {2015})}\BibitemShut {NoStop}%
\bibitem [{\citenamefont {Wang}\ \emph {et~al.}(2015)\citenamefont {Wang},
  \citenamefont {Yan},\ and\ \citenamefont {Zepf}}]{Wang2015}%
  \BibitemOpen
  \bibfield  {author} {\bibinfo {author} {\bibfnamefont {H.~Y.}\ \bibnamefont
  {Wang}}, \bibinfo {author} {\bibfnamefont {X.~Q.}\ \bibnamefont {Yan}}, \
  and\ \bibinfo {author} {\bibfnamefont {M.}~\bibnamefont {Zepf}},\ }\bibfield
  {title} {\enquote {\bibinfo {title} {Signatures of quantum radiation reaction
  in laser-electron-beam collisions},}\ }\href {\doibase 10.1063/1.4929851}
  {\bibfield  {journal} {\bibinfo  {journal} {Phys. Plasmas}\ }\textbf
  {\bibinfo {volume} {22}},\ \bibinfo {pages} {093103} (\bibinfo {year}
  {2015})}\BibitemShut {NoStop}%
\bibitem [{\citenamefont {Li}\ \emph {et~al.}(2020{\natexlab{a}})\citenamefont
  {Li}, \citenamefont {Qiao}, \citenamefont {Liao}, \citenamefont {Wang},
  \citenamefont {Gan}, \citenamefont {Zhou}, \citenamefont {Zhu},\ and\
  \citenamefont {He}}]{Li2020}%
  \BibitemOpen
  \bibfield  {author} {\bibinfo {author} {\bibfnamefont {X.~B.}\ \bibnamefont
  {Li}}, \bibinfo {author} {\bibfnamefont {B.}~\bibnamefont {Qiao}}, \bibinfo
  {author} {\bibfnamefont {Y.~L.}\ \bibnamefont {Liao}}, \bibinfo {author}
  {\bibfnamefont {J.}~\bibnamefont {Wang}}, \bibinfo {author} {\bibfnamefont
  {L.~F.}\ \bibnamefont {Gan}}, \bibinfo {author} {\bibfnamefont {C.~T.}\
  \bibnamefont {Zhou}}, \bibinfo {author} {\bibfnamefont {S.~P.}\ \bibnamefont
  {Zhu}}, \ and\ \bibinfo {author} {\bibfnamefont {X.~T.}\ \bibnamefont {He}},\
  }\bibfield  {title} {\enquote {\bibinfo {title} {Possible signals in
  differentiating the quantum radiation reaction from the classical one},}\
  }\href {\doibase 10.1103/PhysRevA.101.032108} {\bibfield  {journal} {\bibinfo
   {journal} {Phys. Rev. A}\ }\textbf {\bibinfo {volume} {101}},\ \bibinfo
  {pages} {032108} (\bibinfo {year} {2020}{\natexlab{a}})}\BibitemShut
  {NoStop}%
\bibitem [{\citenamefont {Danson1}\ \emph {et~al.}(2019)\citenamefont
  {Danson1}, \citenamefont {Haefner}, \citenamefont {Bromage}, \citenamefont
  {Butcher}, \citenamefont {Chanteloup}, \citenamefont {Chowdhury},
  \citenamefont {Galvanauskas}, \citenamefont {Gizzi}, \citenamefont {Hein},
  \citenamefont {Hillier}, \citenamefont {Hopps}, \citenamefont {Kato},
  \citenamefont {Khazanov}, \citenamefont {Kodama}, \citenamefont {Korn},
  \citenamefont {Li}, \citenamefont {Li}, \citenamefont {Limpert},
  \citenamefont {Ma}, \citenamefont {Nam}, \citenamefont {Neely}, \citenamefont
  {Papadopoulos}, \citenamefont {Penman}, \citenamefont {Qian}, \citenamefont
  {Rocca}, \citenamefont {Shaykin}, \citenamefont {Siders}, \citenamefont
  {Spindloe}, \citenamefont {Szatm´ari}, \citenamefont {Trines}, \citenamefont
  {Zhu}, \citenamefont {Zhu},\ and\ \citenamefont {Zuegel}}]{Danson2019}%
  \BibitemOpen
  \bibfield  {author} {\bibinfo {author} {\bibfnamefont {Colin~N.}\
  \bibnamefont {Danson1}}, \bibinfo {author} {\bibfnamefont {Constantin}\
  \bibnamefont {Haefner}}, \bibinfo {author} {\bibfnamefont {Jake}\
  \bibnamefont {Bromage}}, \bibinfo {author} {\bibfnamefont {Thomas}\
  \bibnamefont {Butcher}}, \bibinfo {author} {\bibfnamefont
  {Jean-Christophe~F.}\ \bibnamefont {Chanteloup}}, \bibinfo {author}
  {\bibfnamefont {Enam~A.}\ \bibnamefont {Chowdhury}}, \bibinfo {author}
  {\bibfnamefont {Almantas}\ \bibnamefont {Galvanauskas}}, \bibinfo {author}
  {\bibfnamefont {Leonida~A.}\ \bibnamefont {Gizzi}}, \bibinfo {author}
  {\bibfnamefont {Joachim}\ \bibnamefont {Hein}}, \bibinfo {author}
  {\bibfnamefont {David~I.}\ \bibnamefont {Hillier}}, \bibinfo {author}
  {\bibfnamefont {Nicholas~W.}\ \bibnamefont {Hopps}}, \bibinfo {author}
  {\bibfnamefont {Yoshiaki}\ \bibnamefont {Kato}}, \bibinfo {author}
  {\bibfnamefont {Efim~A.}\ \bibnamefont {Khazanov}}, \bibinfo {author}
  {\bibfnamefont {Ryosuke}\ \bibnamefont {Kodama}}, \bibinfo {author}
  {\bibfnamefont {Georg}\ \bibnamefont {Korn}}, \bibinfo {author}
  {\bibfnamefont {Ruxin}\ \bibnamefont {Li}}, \bibinfo {author} {\bibfnamefont
  {Yutong}\ \bibnamefont {Li}}, \bibinfo {author} {\bibfnamefont {Jens}\
  \bibnamefont {Limpert}}, \bibinfo {author} {\bibfnamefont {Jingui}\
  \bibnamefont {Ma}}, \bibinfo {author} {\bibfnamefont {Chang~Hee}\
  \bibnamefont {Nam}}, \bibinfo {author} {\bibfnamefont {David}\ \bibnamefont
  {Neely}}, \bibinfo {author} {\bibfnamefont {Dimitrios}\ \bibnamefont
  {Papadopoulos}}, \bibinfo {author} {\bibfnamefont {Rory~R.}\ \bibnamefont
  {Penman}}, \bibinfo {author} {\bibfnamefont {Liejia}\ \bibnamefont {Qian}},
  \bibinfo {author} {\bibfnamefont {Jorge~J.}\ \bibnamefont {Rocca}}, \bibinfo
  {author} {\bibfnamefont {Andrey~A.}\ \bibnamefont {Shaykin}}, \bibinfo
  {author} {\bibfnamefont {Craig~W.}\ \bibnamefont {Siders}}, \bibinfo {author}
  {\bibfnamefont {Christopher}\ \bibnamefont {Spindloe}}, \bibinfo {author}
  {\bibfnamefont {S´andor}\ \bibnamefont {Szatm´ari}}, \bibinfo {author}
  {\bibfnamefont {Raoul M. G.~M.}\ \bibnamefont {Trines}}, \bibinfo {author}
  {\bibfnamefont {Jianqiang}\ \bibnamefont {Zhu}}, \bibinfo {author}
  {\bibfnamefont {Ping}\ \bibnamefont {Zhu}}, \ and\ \bibinfo {author}
  {\bibfnamefont {Jonathan~D.}\ \bibnamefont {Zuegel}},\ }\bibfield  {title}
  {\enquote {\bibinfo {title} {Petawatt and exawatt class lasers worldwide},}\
  }\href@noop {} {\bibfield  {journal} {\bibinfo  {journal} {High Power Laser
  Sci. Eng.}\ }\textbf {\bibinfo {volume} {7}},\ \bibinfo {pages} {e54}
  (\bibinfo {year} {2019})}\BibitemShut {NoStop}%
\bibitem [{\citenamefont {Di~Piazza}\ \emph {et~al.}(2012)\citenamefont
  {Di~Piazza}, \citenamefont {M\"uller}, \citenamefont {Hatsagortsyan},\ and\
  \citenamefont {Keitel}}]{Piazza2012}%
  \BibitemOpen
  \bibfield  {author} {\bibinfo {author} {\bibfnamefont {A.}~\bibnamefont
  {Di~Piazza}}, \bibinfo {author} {\bibfnamefont {C.}~\bibnamefont {M\"uller}},
  \bibinfo {author} {\bibfnamefont {K.~Z.}\ \bibnamefont {Hatsagortsyan}}, \
  and\ \bibinfo {author} {\bibfnamefont {C.~H.}\ \bibnamefont {Keitel}},\
  }\bibfield  {title} {\enquote {\bibinfo {title} {Extremely high-intensity
  laser interactions with fundamental quantum systems},}\ }\href {\doibase
  10.1103/RevModPhys.84.1177} {\bibfield  {journal} {\bibinfo  {journal} {Rev.
  Mod. Phys.}\ }\textbf {\bibinfo {volume} {84}},\ \bibinfo {pages}
  {1177--1228} (\bibinfo {year} {2012})}\BibitemShut {NoStop}%
\bibitem [{\citenamefont {Cole}\ \emph {et~al.}(2018)\citenamefont {Cole},
  \citenamefont {Behm}, \citenamefont {Gerstmayr}, \citenamefont {Blackburn},
  \citenamefont {Wood}, \citenamefont {Baird}, \citenamefont {Duff},
  \citenamefont {Harvey}, \citenamefont {Ilderton}, \citenamefont {Joglekar},
  \citenamefont {Krushelnick}, \citenamefont {Kuschel}, \citenamefont
  {Marklund}, \citenamefont {McKenna}, \citenamefont {Murphy}, \citenamefont
  {Poder}, \citenamefont {Ridgers}, \citenamefont {Samarin}, \citenamefont
  {Sarri}, \citenamefont {Symes}, \citenamefont {Thomas}, \citenamefont
  {Warwick}, \citenamefont {Zepf}, \citenamefont {Najmudin},\ and\
  \citenamefont {Mangles}}]{Cole2018}%
  \BibitemOpen
  \bibfield  {author} {\bibinfo {author} {\bibfnamefont {J.~M.}\ \bibnamefont
  {Cole}}, \bibinfo {author} {\bibfnamefont {K.~T.}\ \bibnamefont {Behm}},
  \bibinfo {author} {\bibfnamefont {E.}~\bibnamefont {Gerstmayr}}, \bibinfo
  {author} {\bibfnamefont {T.~G.}\ \bibnamefont {Blackburn}}, \bibinfo {author}
  {\bibfnamefont {J.~C.}\ \bibnamefont {Wood}}, \bibinfo {author}
  {\bibfnamefont {C.~D.}\ \bibnamefont {Baird}}, \bibinfo {author}
  {\bibfnamefont {M.~J.}\ \bibnamefont {Duff}}, \bibinfo {author}
  {\bibfnamefont {C.}~\bibnamefont {Harvey}}, \bibinfo {author} {\bibfnamefont
  {A.}~\bibnamefont {Ilderton}}, \bibinfo {author} {\bibfnamefont {A.~S.}\
  \bibnamefont {Joglekar}}, \bibinfo {author} {\bibfnamefont {K.}~\bibnamefont
  {Krushelnick}}, \bibinfo {author} {\bibfnamefont {S.}~\bibnamefont
  {Kuschel}}, \bibinfo {author} {\bibfnamefont {M.}~\bibnamefont {Marklund}},
  \bibinfo {author} {\bibfnamefont {P.}~\bibnamefont {McKenna}}, \bibinfo
  {author} {\bibfnamefont {C.~D.}\ \bibnamefont {Murphy}}, \bibinfo {author}
  {\bibfnamefont {K.}~\bibnamefont {Poder}}, \bibinfo {author} {\bibfnamefont
  {C.~P.}\ \bibnamefont {Ridgers}}, \bibinfo {author} {\bibfnamefont {G.~M.}\
  \bibnamefont {Samarin}}, \bibinfo {author} {\bibfnamefont {G.}~\bibnamefont
  {Sarri}}, \bibinfo {author} {\bibfnamefont {D.~R.}\ \bibnamefont {Symes}},
  \bibinfo {author} {\bibfnamefont {A.~G.~R.}\ \bibnamefont {Thomas}}, \bibinfo
  {author} {\bibfnamefont {J.}~\bibnamefont {Warwick}}, \bibinfo {author}
  {\bibfnamefont {M.}~\bibnamefont {Zepf}}, \bibinfo {author} {\bibfnamefont
  {Z.}~\bibnamefont {Najmudin}}, \ and\ \bibinfo {author} {\bibfnamefont
  {S.~P.~D.}\ \bibnamefont {Mangles}},\ }\bibfield  {title} {\enquote {\bibinfo
  {title} {Experimental evidence of radiation reaction in the collision of a
  high-intensity laser pulse with a laser-wakefield accelerated electron
  beam},}\ }\href {\doibase 10.1103/PhysRevX.8.011020} {\bibfield  {journal}
  {\bibinfo  {journal} {Phys. Rev. X}\ }\textbf {\bibinfo {volume} {8}},\
  \bibinfo {pages} {011020} (\bibinfo {year} {2018})}\BibitemShut {NoStop}%
\bibitem [{\citenamefont {Poder}\ \emph {et~al.}(2018)\citenamefont {Poder},
  \citenamefont {Tamburini}, \citenamefont {Sarri}, \citenamefont {Di~Piazza},
  \citenamefont {Kuschel}, \citenamefont {Baird}, \citenamefont {Behm},
  \citenamefont {Bohlen}, \citenamefont {Cole}, \citenamefont {Corvan},
  \citenamefont {Duff}, \citenamefont {Gerstmayr}, \citenamefont {Keitel},
  \citenamefont {Krushelnick}, \citenamefont {Mangles}, \citenamefont
  {McKenna}, \citenamefont {Murphy}, \citenamefont {Najmudin}, \citenamefont
  {Ridgers}, \citenamefont {Samarin}, \citenamefont {Symes}, \citenamefont
  {Thomas}, \citenamefont {Warwick},\ and\ \citenamefont {Zepf}}]{Poder2018}%
  \BibitemOpen
  \bibfield  {author} {\bibinfo {author} {\bibfnamefont {K.}~\bibnamefont
  {Poder}}, \bibinfo {author} {\bibfnamefont {M.}~\bibnamefont {Tamburini}},
  \bibinfo {author} {\bibfnamefont {G.}~\bibnamefont {Sarri}}, \bibinfo
  {author} {\bibfnamefont {A.}~\bibnamefont {Di~Piazza}}, \bibinfo {author}
  {\bibfnamefont {S.}~\bibnamefont {Kuschel}}, \bibinfo {author} {\bibfnamefont
  {C.~D.}\ \bibnamefont {Baird}}, \bibinfo {author} {\bibfnamefont
  {K.}~\bibnamefont {Behm}}, \bibinfo {author} {\bibfnamefont {S.}~\bibnamefont
  {Bohlen}}, \bibinfo {author} {\bibfnamefont {J.~M.}\ \bibnamefont {Cole}},
  \bibinfo {author} {\bibfnamefont {D.~J.}\ \bibnamefont {Corvan}}, \bibinfo
  {author} {\bibfnamefont {M.}~\bibnamefont {Duff}}, \bibinfo {author}
  {\bibfnamefont {E.}~\bibnamefont {Gerstmayr}}, \bibinfo {author}
  {\bibfnamefont {C.~H.}\ \bibnamefont {Keitel}}, \bibinfo {author}
  {\bibfnamefont {K.}~\bibnamefont {Krushelnick}}, \bibinfo {author}
  {\bibfnamefont {S.~P.~D.}\ \bibnamefont {Mangles}}, \bibinfo {author}
  {\bibfnamefont {P.}~\bibnamefont {McKenna}}, \bibinfo {author} {\bibfnamefont
  {C.~D.}\ \bibnamefont {Murphy}}, \bibinfo {author} {\bibfnamefont
  {Z.}~\bibnamefont {Najmudin}}, \bibinfo {author} {\bibfnamefont {C.~P.}\
  \bibnamefont {Ridgers}}, \bibinfo {author} {\bibfnamefont {G.~M.}\
  \bibnamefont {Samarin}}, \bibinfo {author} {\bibfnamefont {D.~R.}\
  \bibnamefont {Symes}}, \bibinfo {author} {\bibfnamefont {A.~G.~R.}\
  \bibnamefont {Thomas}}, \bibinfo {author} {\bibfnamefont {J.}~\bibnamefont
  {Warwick}}, \ and\ \bibinfo {author} {\bibfnamefont {M.}~\bibnamefont
  {Zepf}},\ }\bibfield  {title} {\enquote {\bibinfo {title} {Experimental
  signatures of the quantum nature of radiation reaction in the field of an
  ultraintense laser},}\ }\href {\doibase 10.1103/PhysRevX.8.031004} {\bibfield
   {journal} {\bibinfo  {journal} {Phys. Rev. X}\ }\textbf {\bibinfo {volume}
  {8}},\ \bibinfo {pages} {031004} (\bibinfo {year} {2018})}\BibitemShut
  {NoStop}%
\bibitem [{\citenamefont {Koga}\ \emph {et~al.}(2005)\citenamefont {Koga},
  \citenamefont {Esirkepov},\ and\ \citenamefont {Bulanov}}]{Koga2005}%
  \BibitemOpen
  \bibfield  {author} {\bibinfo {author} {\bibfnamefont {James}\ \bibnamefont
  {Koga}}, \bibinfo {author} {\bibfnamefont {Timur~Zh.}\ \bibnamefont
  {Esirkepov}}, \ and\ \bibinfo {author} {\bibfnamefont {Sergei~V.}\
  \bibnamefont {Bulanov}},\ }\bibfield  {title} {\enquote {\bibinfo {title}
  {Nonlinear thomson scattering in the strong radiation damping regime},}\
  }\href {\doibase 10.1063/1.2013067} {\bibfield  {journal} {\bibinfo
  {journal} {Phys. Plasmas}\ }\textbf {\bibinfo {volume} {12}},\ \bibinfo
  {pages} {093106} (\bibinfo {year} {2005})}\BibitemShut {NoStop}%
\bibitem [{\citenamefont {Ritus}(1985)}]{Ritus1985}%
  \BibitemOpen
  \bibfield  {author} {\bibinfo {author} {\bibfnamefont {V.~I.}\ \bibnamefont
  {Ritus}},\ }\href@noop {} {\bibfield  {journal} {\bibinfo  {journal} {J. Sov.
  Laser Res.}\ }\textbf {\bibinfo {volume} {6}},\ \bibinfo {pages} {497}
  (\bibinfo {year} {1985})}\BibitemShut {NoStop}%
\bibitem [{\citenamefont {Arran}\ \emph {et~al.}(2019)\citenamefont {Arran},
  \citenamefont {Cole}, \citenamefont {Gerstmayr}, \citenamefont {Blackburn},
  \citenamefont {Mangles},\ and\ \citenamefont {Ridgers}}]{Arran2019}%
  \BibitemOpen
  \bibfield  {author} {\bibinfo {author} {\bibfnamefont {C}~\bibnamefont
  {Arran}}, \bibinfo {author} {\bibfnamefont {J~M}\ \bibnamefont {Cole}},
  \bibinfo {author} {\bibfnamefont {E}~\bibnamefont {Gerstmayr}}, \bibinfo
  {author} {\bibfnamefont {T~G}\ \bibnamefont {Blackburn}}, \bibinfo {author}
  {\bibfnamefont {S~P~D}\ \bibnamefont {Mangles}}, \ and\ \bibinfo {author}
  {\bibfnamefont {C~P}\ \bibnamefont {Ridgers}},\ }\bibfield  {title} {\enquote
  {\bibinfo {title} {Optimal parameters for radiation reaction experiments},}\
  }\href {\doibase 10.1088/1361-6587/ab20f6} {\bibfield  {journal} {\bibinfo
  {journal} {Plasma Phys. Contr. F.}\ }\textbf {\bibinfo {volume} {61}},\
  \bibinfo {pages} {074009} (\bibinfo {year} {2019})}\BibitemShut {NoStop}%
\bibitem [{\citenamefont {Khokonov}\ and\ \citenamefont
  {Nitta}(2002)}]{Khokonov2002}%
  \BibitemOpen
  \bibfield  {author} {\bibinfo {author} {\bibfnamefont {M.~Kh.}\ \bibnamefont
  {Khokonov}}\ and\ \bibinfo {author} {\bibfnamefont {H.}~\bibnamefont
  {Nitta}},\ }\bibfield  {title} {\enquote {\bibinfo {title} {Standard
  radiation spectrum of relativistic electrons: Beyond the synchrotron
  approximation},}\ }\href {\doibase 10.1103/PhysRevLett.89.094801} {\bibfield
  {journal} {\bibinfo  {journal} {Phys. Rev. Lett.}\ }\textbf {\bibinfo
  {volume} {89}},\ \bibinfo {pages} {094801} (\bibinfo {year}
  {2002})}\BibitemShut {NoStop}%
\bibitem [{\citenamefont {Di~Piazza}\ \emph {et~al.}(2018)\citenamefont
  {Di~Piazza}, \citenamefont {Tamburini}, \citenamefont {Meuren},\ and\
  \citenamefont {Keitel}}]{Piazza2018}%
  \BibitemOpen
  \bibfield  {author} {\bibinfo {author} {\bibfnamefont {A.}~\bibnamefont
  {Di~Piazza}}, \bibinfo {author} {\bibfnamefont {M.}~\bibnamefont
  {Tamburini}}, \bibinfo {author} {\bibfnamefont {S.}~\bibnamefont {Meuren}}, \
  and\ \bibinfo {author} {\bibfnamefont {C.~H.}\ \bibnamefont {Keitel}},\
  }\bibfield  {title} {\enquote {\bibinfo {title} {Implementing nonlinear
  compton scattering beyond the local-constant-field approximation},}\ }\href
  {\doibase 10.1103/PhysRevA.98.012134} {\bibfield  {journal} {\bibinfo
  {journal} {Phys. Rev. A}\ }\textbf {\bibinfo {volume} {98}},\ \bibinfo
  {pages} {012134} (\bibinfo {year} {2018})}\BibitemShut {NoStop}%
\bibitem [{\citenamefont {Di~Piazza}\ \emph {et~al.}(2019)\citenamefont
  {Di~Piazza}, \citenamefont {Tamburini}, \citenamefont {Meuren},\ and\
  \citenamefont {Keitel}}]{Piazza2019}%
  \BibitemOpen
  \bibfield  {author} {\bibinfo {author} {\bibfnamefont {A.}~\bibnamefont
  {Di~Piazza}}, \bibinfo {author} {\bibfnamefont {M.}~\bibnamefont
  {Tamburini}}, \bibinfo {author} {\bibfnamefont {S.}~\bibnamefont {Meuren}}, \
  and\ \bibinfo {author} {\bibfnamefont {C.~H.}\ \bibnamefont {Keitel}},\
  }\bibfield  {title} {\enquote {\bibinfo {title} {Improved
  local-constant-field approximation for strong-field qed codes},}\ }\href
  {\doibase 10.1103/PhysRevA.99.022125} {\bibfield  {journal} {\bibinfo
  {journal} {Phys. Rev. A}\ }\textbf {\bibinfo {volume} {99}},\ \bibinfo
  {pages} {022125} (\bibinfo {year} {2019})}\BibitemShut {NoStop}%
\bibitem [{\citenamefont {Harvey}\ \emph {et~al.}(2015)\citenamefont {Harvey},
  \citenamefont {Ilderton},\ and\ \citenamefont {King}}]{Harvey2015}%
  \BibitemOpen
  \bibfield  {author} {\bibinfo {author} {\bibfnamefont {C.~N.}\ \bibnamefont
  {Harvey}}, \bibinfo {author} {\bibfnamefont {A.}~\bibnamefont {Ilderton}}, \
  and\ \bibinfo {author} {\bibfnamefont {B.}~\bibnamefont {King}},\ }\bibfield
  {title} {\enquote {\bibinfo {title} {Testing numerical implementations of
  strong-field electrodynamics},}\ }\href {\doibase 10.1103/PhysRevA.91.013822}
  {\bibfield  {journal} {\bibinfo  {journal} {Phys. Rev. A}\ }\textbf {\bibinfo
  {volume} {91}},\ \bibinfo {pages} {013822} (\bibinfo {year}
  {2015})}\BibitemShut {NoStop}%
\bibitem [{\citenamefont {Green}\ and\ \citenamefont
  {Harvey}(2015)}]{Green2015}%
  \BibitemOpen
  \bibfield  {author} {\bibinfo {author} {\bibfnamefont {D.G.}\ \bibnamefont
  {Green}}\ and\ \bibinfo {author} {\bibfnamefont {C.N.}\ \bibnamefont
  {Harvey}},\ }\bibfield  {title} {\enquote {\bibinfo {title} {Simla:
  Simulating particle dynamics in intense laser and other electromagnetic
  fields via classical and quantum electrodynamics},}\ }\href {\doibase
  https://doi.org/10.1016/j.cpc.2015.02.030} {\bibfield  {journal} {\bibinfo
  {journal} {Comput. Phys. Commun.}\ }\textbf {\bibinfo {volume} {192}},\
  \bibinfo {pages} {313 -- 321} (\bibinfo {year} {2015})}\BibitemShut {NoStop}%
\bibitem [{\citenamefont {Elkina}\ \emph {et~al.}(2011)\citenamefont {Elkina},
  \citenamefont {Fedotov}, \citenamefont {Kostyukov}, \citenamefont {Legkov},
  \citenamefont {Narozhny}, \citenamefont {Nerush},\ and\ \citenamefont
  {Ruhl}}]{Elkina2011}%
  \BibitemOpen
  \bibfield  {author} {\bibinfo {author} {\bibfnamefont {N.~V.}\ \bibnamefont
  {Elkina}}, \bibinfo {author} {\bibfnamefont {A.~M.}\ \bibnamefont {Fedotov}},
  \bibinfo {author} {\bibfnamefont {I.~Yu.}\ \bibnamefont {Kostyukov}},
  \bibinfo {author} {\bibfnamefont {M.~V.}\ \bibnamefont {Legkov}}, \bibinfo
  {author} {\bibfnamefont {N.~B.}\ \bibnamefont {Narozhny}}, \bibinfo {author}
  {\bibfnamefont {E.~N.}\ \bibnamefont {Nerush}}, \ and\ \bibinfo {author}
  {\bibfnamefont {H.}~\bibnamefont {Ruhl}},\ }\bibfield  {title} {\enquote
  {\bibinfo {title} {Qed cascades induced by circularly polarized laser
  fields},}\ }\href {\doibase 10.1103/PhysRevSTAB.14.054401} {\bibfield
  {journal} {\bibinfo  {journal} {Phys. Rev. ST Accel. Beams}\ }\textbf
  {\bibinfo {volume} {14}},\ \bibinfo {pages} {054401} (\bibinfo {year}
  {2011})}\BibitemShut {NoStop}%
\bibitem [{\citenamefont {Ridgers}\ \emph {et~al.}(2014)\citenamefont
  {Ridgers}, \citenamefont {Kirk}, \citenamefont {Duclous}, \citenamefont
  {Blackburn}, \citenamefont {Brady}, \citenamefont {Bennett}, \citenamefont
  {Arber},\ and\ \citenamefont {Bell}}]{Ridgers2014}%
  \BibitemOpen
  \bibfield  {author} {\bibinfo {author} {\bibfnamefont {C.P.}\ \bibnamefont
  {Ridgers}}, \bibinfo {author} {\bibfnamefont {J.G.}\ \bibnamefont {Kirk}},
  \bibinfo {author} {\bibfnamefont {R.}~\bibnamefont {Duclous}}, \bibinfo
  {author} {\bibfnamefont {T.G.}\ \bibnamefont {Blackburn}}, \bibinfo {author}
  {\bibfnamefont {C.S.}\ \bibnamefont {Brady}}, \bibinfo {author}
  {\bibfnamefont {K.}~\bibnamefont {Bennett}}, \bibinfo {author} {\bibfnamefont
  {T.D.}\ \bibnamefont {Arber}}, \ and\ \bibinfo {author} {\bibfnamefont
  {A.R.}\ \bibnamefont {Bell}},\ }\bibfield  {title} {\enquote {\bibinfo
  {title} {Modelling gamma-ray photon emission and pair production in
  high-intensity laser–matter interactions},}\ }\href {\doibase
  https://doi.org/10.1016/j.jcp.2013.12.007} {\bibfield  {journal} {\bibinfo
  {journal} {J. Comput. Phys.}\ }\textbf {\bibinfo {volume} {260}},\ \bibinfo
  {pages} {273 -- 285} (\bibinfo {year} {2014})}\BibitemShut {NoStop}%
\bibitem [{\citenamefont {Baier}\ \emph {et~al.}(1998)\citenamefont {Baier},
  \citenamefont {Katkov},\ and\ \citenamefont {Strakhovenko}}]{Baier1998}%
  \BibitemOpen
  \bibfield  {author} {\bibinfo {author} {\bibfnamefont {V.~N.}\ \bibnamefont
  {Baier}}, \bibinfo {author} {\bibfnamefont {V.~M.}\ \bibnamefont {Katkov}}, \
  and\ \bibinfo {author} {\bibfnamefont {V.~M.}\ \bibnamefont {Strakhovenko}},\
  }\href@noop {} {\emph {\bibinfo {title} {Electromagnetic Processes at High
  Energies in Oriented Single Crystals}}}\ (\bibinfo  {publisher} {World
  Scientific},\ \bibinfo {address} {Singapore},\ \bibinfo {year}
  {1998})\BibitemShut {NoStop}%
\bibitem [{\citenamefont {Gonoskov}\ \emph {et~al.}(2015)\citenamefont
  {Gonoskov}, \citenamefont {Bastrakov}, \citenamefont {Efimenko},
  \citenamefont {Ilderton}, \citenamefont {Marklund}, \citenamefont {Meyerov},
  \citenamefont {Muraviev}, \citenamefont {Sergeev}, \citenamefont {Surmin},\
  and\ \citenamefont {Wallin}}]{Gonoskov2015}%
  \BibitemOpen
  \bibfield  {author} {\bibinfo {author} {\bibfnamefont {A.}~\bibnamefont
  {Gonoskov}}, \bibinfo {author} {\bibfnamefont {S.}~\bibnamefont {Bastrakov}},
  \bibinfo {author} {\bibfnamefont {E.}~\bibnamefont {Efimenko}}, \bibinfo
  {author} {\bibfnamefont {A.}~\bibnamefont {Ilderton}}, \bibinfo {author}
  {\bibfnamefont {M.}~\bibnamefont {Marklund}}, \bibinfo {author}
  {\bibfnamefont {I.}~\bibnamefont {Meyerov}}, \bibinfo {author} {\bibfnamefont
  {A.}~\bibnamefont {Muraviev}}, \bibinfo {author} {\bibfnamefont
  {A.}~\bibnamefont {Sergeev}}, \bibinfo {author} {\bibfnamefont
  {I.}~\bibnamefont {Surmin}}, \ and\ \bibinfo {author} {\bibfnamefont
  {E.}~\bibnamefont {Wallin}},\ }\bibfield  {title} {\enquote {\bibinfo {title}
  {Extended particle-in-cell schemes for physics in ultrastrong laser fields:
  Review and developments},}\ }\href {\doibase 10.1103/PhysRevE.92.023305}
  {\bibfield  {journal} {\bibinfo  {journal} {Phys. Rev. E}\ }\textbf {\bibinfo
  {volume} {92}},\ \bibinfo {pages} {023305} (\bibinfo {year}
  {2015})}\BibitemShut {NoStop}%
\bibitem [{\citenamefont {Li}\ \emph {et~al.}(2019)\citenamefont {Li},
  \citenamefont {Shaisultanov}, \citenamefont {Hatsagortsyan}, \citenamefont
  {Wan}, \citenamefont {Keitel},\ and\ \citenamefont {Li}}]{Liyf2019}%
  \BibitemOpen
  \bibfield  {author} {\bibinfo {author} {\bibfnamefont {Yan-Fei}\ \bibnamefont
  {Li}}, \bibinfo {author} {\bibfnamefont {Rashid}\ \bibnamefont
  {Shaisultanov}}, \bibinfo {author} {\bibfnamefont {Karen~Z.}\ \bibnamefont
  {Hatsagortsyan}}, \bibinfo {author} {\bibfnamefont {Feng}\ \bibnamefont
  {Wan}}, \bibinfo {author} {\bibfnamefont {Christoph~H.}\ \bibnamefont
  {Keitel}}, \ and\ \bibinfo {author} {\bibfnamefont {Jian-Xing}\ \bibnamefont
  {Li}},\ }\bibfield  {title} {\enquote {\bibinfo {title} {Ultrarelativistic
  electron-beam polarization in single-shot interaction with an ultraintense
  laser pulse},}\ }\href {\doibase 10.1103/PhysRevLett.122.154801} {\bibfield
  {journal} {\bibinfo  {journal} {Phys. Rev. Lett.}\ }\textbf {\bibinfo
  {volume} {122}},\ \bibinfo {pages} {154801} (\bibinfo {year}
  {2019})}\BibitemShut {NoStop}%
\bibitem [{\citenamefont {Li}\ \emph {et~al.}(2020{\natexlab{b}})\citenamefont
  {Li}, \citenamefont {Shaisultanov}, \citenamefont {Chen}, \citenamefont
  {Wan}, \citenamefont {Hatsagortsyan}, \citenamefont {Keitel},\ and\
  \citenamefont {Li}}]{Liyf2020}%
  \BibitemOpen
  \bibfield  {author} {\bibinfo {author} {\bibfnamefont {Yan-Fei}\ \bibnamefont
  {Li}}, \bibinfo {author} {\bibfnamefont {Rashid}\ \bibnamefont
  {Shaisultanov}}, \bibinfo {author} {\bibfnamefont {Yue-Yue}\ \bibnamefont
  {Chen}}, \bibinfo {author} {\bibfnamefont {Feng}\ \bibnamefont {Wan}},
  \bibinfo {author} {\bibfnamefont {Karen~Z.}\ \bibnamefont {Hatsagortsyan}},
  \bibinfo {author} {\bibfnamefont {Christoph~H.}\ \bibnamefont {Keitel}}, \
  and\ \bibinfo {author} {\bibfnamefont {Jian-Xing}\ \bibnamefont {Li}},\
  }\bibfield  {title} {\enquote {\bibinfo {title} {Polarized ultrashort
  brilliant multi-gev $\ensuremath{\gamma}$ rays via single-shot laser-electron
  interaction},}\ }\href {\doibase 10.1103/PhysRevLett.124.014801} {\bibfield
  {journal} {\bibinfo  {journal} {Phys. Rev. Lett.}\ }\textbf {\bibinfo
  {volume} {124}},\ \bibinfo {pages} {014801} (\bibinfo {year}
  {2020}{\natexlab{b}})}\BibitemShut {NoStop}%
\bibitem [{\citenamefont {Burton}\ and\ \citenamefont
  {Noble}(2014)}]{Burton2014}%
  \BibitemOpen
  \bibfield  {author} {\bibinfo {author} {\bibfnamefont {David~A.}\
  \bibnamefont {Burton}}\ and\ \bibinfo {author} {\bibfnamefont {Adam}\
  \bibnamefont {Noble}},\ }\bibfield  {title} {\enquote {\bibinfo {title}
  {Aspects of electromagnetic radiation reaction in strong fields},}\ }\href
  {\doibase 10.1080/00107514.2014.886840} {\bibfield  {journal} {\bibinfo
  {journal} {Contemporary Physics}\ }\textbf {\bibinfo {volume} {55}},\
  \bibinfo {pages} {110--121} (\bibinfo {year} {2014})}\BibitemShut {NoStop}%
\bibitem [{\citenamefont {Vranic}\ \emph {et~al.}(2016)\citenamefont {Vranic},
  \citenamefont {Martins}, \citenamefont {Fonseca},\ and\ \citenamefont
  {Silva}}]{Vranic2016}%
  \BibitemOpen
  \bibfield  {author} {\bibinfo {author} {\bibfnamefont {M.}~\bibnamefont
  {Vranic}}, \bibinfo {author} {\bibfnamefont {J.L.}\ \bibnamefont {Martins}},
  \bibinfo {author} {\bibfnamefont {R.A.}\ \bibnamefont {Fonseca}}, \ and\
  \bibinfo {author} {\bibfnamefont {L.O.}\ \bibnamefont {Silva}},\ }\bibfield
  {title} {\enquote {\bibinfo {title} {Classical radiation reaction in
  particle-in-cell simulations},}\ }\href {\doibase
  https://doi.org/10.1016/j.cpc.2016.04.002} {\bibfield  {journal} {\bibinfo
  {journal} {Comput. Phys. Commun.}\ }\textbf {\bibinfo {volume} {204}},\
  \bibinfo {pages} {141 -- 151} (\bibinfo {year} {2016})}\BibitemShut {NoStop}%
\bibitem [{\citenamefont {of~CAIN Version~2.42}()}]{CAIN}%
  \BibitemOpen
  \bibfield  {author} {\bibinfo {author} {\bibfnamefont {User’s~Manual}\
  \bibnamefont {of~CAIN Version~2.42}},\ }\href@noop {} {\ }\bibinfo {note}
  {\url{ http://lcdev.kek.jp/~yokoya/CAIN/}}\BibitemShut {NoStop}%
\bibitem [{\citenamefont {Esarey}\ \emph {et~al.}(2009)\citenamefont {Esarey},
  \citenamefont {Schroeder},\ and\ \citenamefont {Leemans}}]{Esarey2009}%
  \BibitemOpen
  \bibfield  {author} {\bibinfo {author} {\bibfnamefont {E.}~\bibnamefont
  {Esarey}}, \bibinfo {author} {\bibfnamefont {C.~B.}\ \bibnamefont
  {Schroeder}}, \ and\ \bibinfo {author} {\bibfnamefont {W.~P.}\ \bibnamefont
  {Leemans}},\ }\bibfield  {title} {\enquote {\bibinfo {title} {Physics of
  laser-driven plasma-based electron accelerators},}\ }\href {\doibase
  10.1103/RevModPhys.81.1229} {\bibfield  {journal} {\bibinfo  {journal} {Rev.
  Mod. Phys.}\ }\textbf {\bibinfo {volume} {81}},\ \bibinfo {pages}
  {1229--1285} (\bibinfo {year} {2009})}\BibitemShut {NoStop}%
\bibitem [{\citenamefont {Leemans}\ \emph {et~al.}(2014)\citenamefont
  {Leemans}, \citenamefont {Gonsalves}, \citenamefont {Mao}, \citenamefont
  {Nakamura}, \citenamefont {Benedetti}, \citenamefont {Schroeder},
  \citenamefont {T\'oth}, \citenamefont {Daniels}, \citenamefont
  {Mittelberger}, \citenamefont {Bulanov}, \citenamefont {Vay}, \citenamefont
  {Geddes},\ and\ \citenamefont {Esarey}}]{Leemans2014}%
  \BibitemOpen
  \bibfield  {author} {\bibinfo {author} {\bibfnamefont {W.~P.}\ \bibnamefont
  {Leemans}}, \bibinfo {author} {\bibfnamefont {A.~J.}\ \bibnamefont
  {Gonsalves}}, \bibinfo {author} {\bibfnamefont {H.-S.}\ \bibnamefont {Mao}},
  \bibinfo {author} {\bibfnamefont {K.}~\bibnamefont {Nakamura}}, \bibinfo
  {author} {\bibfnamefont {C.}~\bibnamefont {Benedetti}}, \bibinfo {author}
  {\bibfnamefont {C.~B.}\ \bibnamefont {Schroeder}}, \bibinfo {author}
  {\bibfnamefont {Cs.}\ \bibnamefont {T\'oth}}, \bibinfo {author}
  {\bibfnamefont {J.}~\bibnamefont {Daniels}}, \bibinfo {author} {\bibfnamefont
  {D.~E.}\ \bibnamefont {Mittelberger}}, \bibinfo {author} {\bibfnamefont
  {S.~S.}\ \bibnamefont {Bulanov}}, \bibinfo {author} {\bibfnamefont {J.-L.}\
  \bibnamefont {Vay}}, \bibinfo {author} {\bibfnamefont {C.~G.~R.}\
  \bibnamefont {Geddes}}, \ and\ \bibinfo {author} {\bibfnamefont
  {E.}~\bibnamefont {Esarey}},\ }\bibfield  {title} {\enquote {\bibinfo {title}
  {Multi-gev electron beams from capillary-discharge-guided subpetawatt laser
  pulses in the self-trapping regime},}\ }\href {\doibase
  10.1103/PhysRevLett.113.245002} {\bibfield  {journal} {\bibinfo  {journal}
  {Phys. Rev. Lett.}\ }\textbf {\bibinfo {volume} {113}},\ \bibinfo {pages}
  {245002} (\bibinfo {year} {2014})}\BibitemShut {NoStop}%
\bibitem [{\citenamefont {Gonsalves}\ \emph {et~al.}(2019)\citenamefont
  {Gonsalves}, \citenamefont {Nakamura}, \citenamefont {Daniels}, \citenamefont
  {Benedetti}, \citenamefont {Pieronek}, \citenamefont {de~Raadt},
  \citenamefont {Steinke}, \citenamefont {Bin}, \citenamefont {Bulanov},
  \citenamefont {van Tilborg}, \citenamefont {Geddes}, \citenamefont
  {Schroeder}, \citenamefont {T\'oth}, \citenamefont {Esarey}, \citenamefont
  {Swanson}, \citenamefont {Fan-Chiang}, \citenamefont {Bagdasarov},
  \citenamefont {Bobrova}, \citenamefont {Gasilov}, \citenamefont {Korn},
  \citenamefont {Sasorov},\ and\ \citenamefont {Leemans}}]{Gonsalves2019}%
  \BibitemOpen
  \bibfield  {author} {\bibinfo {author} {\bibfnamefont {A.~J.}\ \bibnamefont
  {Gonsalves}}, \bibinfo {author} {\bibfnamefont {K.}~\bibnamefont {Nakamura}},
  \bibinfo {author} {\bibfnamefont {J.}~\bibnamefont {Daniels}}, \bibinfo
  {author} {\bibfnamefont {C.}~\bibnamefont {Benedetti}}, \bibinfo {author}
  {\bibfnamefont {C.}~\bibnamefont {Pieronek}}, \bibinfo {author}
  {\bibfnamefont {T.~C.~H.}\ \bibnamefont {de~Raadt}}, \bibinfo {author}
  {\bibfnamefont {S.}~\bibnamefont {Steinke}}, \bibinfo {author} {\bibfnamefont
  {J.~H.}\ \bibnamefont {Bin}}, \bibinfo {author} {\bibfnamefont {S.~S.}\
  \bibnamefont {Bulanov}}, \bibinfo {author} {\bibfnamefont {J.}~\bibnamefont
  {van Tilborg}}, \bibinfo {author} {\bibfnamefont {C.~G.~R.}\ \bibnamefont
  {Geddes}}, \bibinfo {author} {\bibfnamefont {C.~B.}\ \bibnamefont
  {Schroeder}}, \bibinfo {author} {\bibfnamefont {Cs.}\ \bibnamefont {T\'oth}},
  \bibinfo {author} {\bibfnamefont {E.}~\bibnamefont {Esarey}}, \bibinfo
  {author} {\bibfnamefont {K.}~\bibnamefont {Swanson}}, \bibinfo {author}
  {\bibfnamefont {L.}~\bibnamefont {Fan-Chiang}}, \bibinfo {author}
  {\bibfnamefont {G.}~\bibnamefont {Bagdasarov}}, \bibinfo {author}
  {\bibfnamefont {N.}~\bibnamefont {Bobrova}}, \bibinfo {author} {\bibfnamefont
  {V.}~\bibnamefont {Gasilov}}, \bibinfo {author} {\bibfnamefont
  {G.}~\bibnamefont {Korn}}, \bibinfo {author} {\bibfnamefont {P.}~\bibnamefont
  {Sasorov}}, \ and\ \bibinfo {author} {\bibfnamefont {W.~P.}\ \bibnamefont
  {Leemans}},\ }\bibfield  {title} {\enquote {\bibinfo {title} {Petawatt laser
  guiding and electron beam acceleration to 8 gev in a laser-heated capillary
  discharge waveguide},}\ }\href {\doibase 10.1103/PhysRevLett.122.084801}
  {\bibfield  {journal} {\bibinfo  {journal} {Phys. Rev. Lett.}\ }\textbf
  {\bibinfo {volume} {122}},\ \bibinfo {pages} {084801} (\bibinfo {year}
  {2019})}\BibitemShut {NoStop}%
\bibitem [{\citenamefont {Quesnel}\ and\ \citenamefont
  {Mora}(1998)}]{Quesnel1998}%
  \BibitemOpen
  \bibfield  {author} {\bibinfo {author} {\bibfnamefont {Brice}\ \bibnamefont
  {Quesnel}}\ and\ \bibinfo {author} {\bibfnamefont {Patrick}\ \bibnamefont
  {Mora}},\ }\bibfield  {title} {\enquote {\bibinfo {title} {Theory and
  simulation of the interaction of ultraintense laser pulses with electrons in
  vacuum},}\ }\href {\doibase 10.1103/PhysRevE.58.3719} {\bibfield  {journal}
  {\bibinfo  {journal} {Phys. Rev. E}\ }\textbf {\bibinfo {volume} {58}},\
  \bibinfo {pages} {3719--3732} (\bibinfo {year} {1998})}\BibitemShut {NoStop}%
\end{thebibliography}%

\end{document}